\documentclass[prd,aps,twocolumn,a4paper20170207,floatfix,nofootinbib]{revtex4}

 \usepackage{graphicx,psfrag}
\usepackage{mathrsfs}
\usepackage{amsmath,amsfonts,amssymb}
\usepackage{multirow,hyperref}
\usepackage{comment}
\usepackage{enumerate}
\usepackage{mathrsfs}
\usepackage{pifont}
\usepackage[normalem]{ulem}
\usepackage{xspace}

\newcommand{\be}{\begin{equation}}
\newcommand{\ee}{\end{equation}}
\newcommand{\bea}{\begin{eqnarray}}
\newcommand{\eea}{\end{eqnarray}}
\newcommand{\bel}{\begin{align}}
\newcommand{\eel}{\end{align}}

\def\GMc2{{\rm G M_{\odot} c^{-2}}}

\def\EOBpp{EOB$_{\rm BBH}$\xspace}
\def\EOBad{EOB$_{\rm adPN}$\xspace}
\def\EOBre{EOB$_{\rm adGSF^+}$\xspace}
\def\EOBdy{EOB$_{\rm dyPN}$\xspace}
\def\EOBPaul{EOB$_{\rm dyGSF^+}$\xspace}

\usepackage{color}

\definecolor{cyan}{rgb}{0,0.9,0.9}
\definecolor{orange}{rgb}{0.9,0.5,0}
\definecolor{magenta}{rgb}{1,0,1}
\definecolor{purple}{rgb}{0.8,0.4,0.8}
\definecolor{gray}{rgb}{0.8242,0.8242,0.8242}

\begin{document}


\title{Comprehensive numerical relativity -- effective-one-body comparison 
       to inform improvements in waveform models for binary neutron star systems}

\author{Tim \surname{Dietrich}}
\author{Tanja \surname{Hinderer}}

 \affiliation{Max Planck Institute for Gravitational Physics 
                    (Albert Einstein Institute), Am M\"uhlenberg 1, Potsdam 14476, Germany}
\date{\today}

\begin{abstract}
We present a detailed comparison between tidal 
effective-one-body (EOB) models 
and new state-of-the-art numerical relativity simulations for non-spinning binary neutron star systems. 
This comparison is the most extensive one to date, covering a wide range in the 
parameter space and encompassing the energetics of the binary, 
the periastron advance, the time and frequency evolution of the gravitational wave phase for the dominant mode, and 
several subdominant modes. We consider different EOB models with tidal effects that have been proposed, 
including the model with dynamical tides of~[Phys.Rev.Lett.~116 (2016) no.18, 181101] 
and the gravitational self-force (GSF) inspired tidal EOB model of~[Phys.Rev.Lett.~114 (2015) no.16, 161103]. 
The EOB model with dynamical tides leads to the best representation 
of the systems considered here, however, 
the differences to the GSF-inspired model are small. 
A common feature is that for systems where matter effects are large, 
i.e.~stiff equations of state or small total masses, all EOB models underestimate the tidal effects and differences to 
the results from numerical relativity simulations become noticeable near the merger. 
We analyze this regime to diagnose the shortcomings of the models in the late inspiral, 
where the two neutron stars are no longer isolated bodies moving in vacuum. 
Our work will serve to guide further advances in modeling these systems.

\end{abstract}

\pacs{
  04.25.D-,     
  04.30.Db,   
  95.30.Sf,     
  95.30.Lz,   
  97.60.Jd      
}

\maketitle

\section{Introduction}
\label{sec:intro}

The recent first detection of gravitational waves (GWs) of 
merging binary black holes (BBHs)~\cite{Abbott:2016blz,Abbott:2016nmj} 
has initiated a new observational era in astronomy. 
Within the next few years GW detections of neutron star (NS) binaries
are also anticipated~\cite{Aasi:2013wya,Abbott:2016ymx}. 
These observations will have a unique potential to probe the equation of state (EOS) of 
the ultradense matter in NS 
interiors~\cite{Read:2009yp,Damour:2012yf,Read:2013zra,DelPozzo:2013ala,Lackey:2014fwa,
Flanagan:2007ix,Hinderer:2007mb,Damour:2009vw,Binnington:2009bb,Hinderer:2009ca}, which remains one of the 
major unsolved problems in modern astrophysics.  

To successfully extract the EOS information from the GW data requires
highly accurate waveform models (templates) in order 
to perform matched-filtered searches, 
where the data are cross-correlated with a template bank covering 
all physical values of the parameters. 
Constructing sufficiently accurate models requires a detailed
understanding of late stages of the binary NS (BNS) coalescence, 
where matter effects on the GW signal become most prominent.
In this regime, one must solve Einstein's equations together with the 
equations of general relativistic hydrodynamics, a challenging task that 
can only be accomplished through numerical relativity (NR) simulations. 
Over the last few years the physical realism of the NR
simulations of BNSs has been significantly advanced by including a refined treatment of 
magnetic fields and microphysical processes, 
see e.g.~\cite{Giacomazzo:2010bx,Sekiguchi:2010fh,Sekiguchi:2011zd,Galeazzi:2013mia,
Kiuchi:2014hja,Dionysopoulou:2015tda,Palenzuela:2015dqa,Lehner:2016lxy,
Sekiguchi:2015dma,Foucart:2015gaa,Sekiguchi:2016bjd}, 
and by improving the numerical accuracy of the predicted GW 
signals~\cite{Bernuzzi:2011aq,Radice:2013hxh,Radice:2013xpa,Hotokezaka:2013mm,
Radice:2015nva,Hotokezaka:2015xka,Bernuzzi:2016pie}. 

Unfortunately, NR simulations are too expensive 
to create template banks for BNS GW signals.
Furthermore, analytical descriptions such as the Post-Newtonian (PN) 
approach become inaccurate close to the merger 
and are therefore inadequate
for analyzing the regime where tidal effects 
are largest~\cite{Favata:2013rwa,Yagi:2013baa,Lackey:2014fwa}. 
A successful way to recast PN results and combine them with information from NR simulations 
is the effective-one-body (EOB) formalism~\cite{Buonanno:1998gg,Buonanno:2000ef}. This framework 
has been refined to devise an accurate, semi-analytical description 
of the dynamics and GW signals of coalescing BHs with arbitrary spins and mass ratios~\cite{Taracchini:2013rva,Pan:2013rra,Damour:2014sva}. 
Tidal effects have also been included in the EOB model, 
starting with the first analysis~\cite{Damour:2009wj} that used Newtonian 
and partial 1PN tidal information~\cite{Vines:2010ca} and
was compared against NR simulations in~\cite{Baiotti:2010xh}.
Further theoretical developments including 2PN tidal effects 
in the conservative dynamics~\cite{Bini:2012gu} 
and 1PN tidal corrections to the waveform amplitudes~\cite{Damour:2012yf} have
led to an improved tidal EOB model that was tested against 
results from improved NR simulations with robust error estimates 
in~\cite{Bernuzzi:2012ci,Hotokezaka:2013mm}. More recent work has focused on further advancing the EOB description of 
tidal effects in different ways. The work of Ref.~\cite{Bini:2014zxa} 
suggested an EOB model based on the Schwarzschild tidal field 
together with gravitational self-force (GSF) corrections. 
The pure GSF-based model underestimates matter effects in the late inspiral, 
but was augmented by an additional term to develop a model 
that remains close to NR predictions~\cite{Bernuzzi:2014owa,Hotokezaka:2015xka}. 
Another line of recent work has focused on the enhancement of 
tidal effects due to the NS's finite fundamental ($f-$) 
mode frequency that also leads to an improved performance of the 
model in comparisons with NR results~\cite{Hinderer:2016eia,Steinhoff:2016rfi}.

To date, EOB -- NR comparisons for BNSs have been limited to equal mass systems. 
Although nearly equal mass systems seem 
to be common~\cite{Swiggum:2015yras,Lattimer:2012nd},
recent observations have also discovered systems with mass 
ratios of $q=M^A/M^B\sim 1.3$~\cite{Martinez:2015mya,Lazarus:2016hfu}. 
The range of pulsar masses that have been accurately 
determined ($1.17-2.01 M_\odot$) and 
`in situ' models, see e.g.~\cite{Dominik:2012kk,Dietrich:2015pxa},
also suggest a larger range of the mass ratio. 
Therefore, in this paper20170207, we go beyond previous work in covering the BNS parameter 
space by considering mass ratios in the range $q=1.0-1.5$. 
Since for unequal-mass systems spherical harmonic modes in the GWs 
besides the dominant $(\ell,m)=(2,2)$ mode are larger than for equal masses, we also 
compare the EOB predictions for the
(2,1), (3,3), and (4,4) modes with NR simulations. 
In addition, we perform a comprehensive study of other important diagnostic 
quantities to test the performance of the EOB models and gain additional insights. 
Specifically, we investigate the phase evolution of the (2,2) mode as a function of the 
GW frequency, analyze the periastron advance (PA), and study the dynamics of the binary 
by tracing the binding energy vs.~specific orbital angular momentum curve over the inspiral. 

In general, as also found in previous EOB-NR comparisons, for most scenarios 
tidal effects in NR simulations are larger than those predicted by EOB models. 
The origin of this trend is twofold. On the one hand, 
tidal EOB models generally underestimate tidal effects present in the last stages of 
the BNS coalescence because they lack higher PN order tidal terms 
and physical effects that become relevant once the 
NSs come into contact shortly before the merger. 
On the other hand NR simulations tend to 
overestimate tidal effects since numerical dissipation 
accelerates the inspiral similar to matter effects. 
This shows that a reliable error estimate is crucial for a proper 
numerical relativity- analytical relativity comparison of BNSs.

The paper20170207 is organized as follows. 
In Sec.~\ref{sec:tidalEOB} we briefly review the
tidal EOB models used in this article. 
Sec.~\ref{sec:config} describes the 
configurations we investigate (see in particular Tab.~\ref{tab:config}) 
and briefly summarizes the NR methods employed in the simulations. 
We then discuss quantities characterizing the dynamics by considering 
binding energy vs.~angular momentum curves in Sec.~\ref{sec:ej} 
and the PA in Sec.~\ref{sec:PA}.
In Sec.~\ref{sec:GWs} we compare the EOB and NR waveforms.  
Finally, we discuss current problems of the EOB models during the last stages 
of the inspiral for two particular cases and
outline aspects requiring further improvements to the EOB model in Sec.~\ref{sec:case}, where we also 
show results from combining the models of~\cite{Bernuzzi:2014owa} 
and~\cite{Hinderer:2016eia} as a possibility to further enhance 
tidal effects in the late inspiral regime. 
We conclude in Sec.~\ref{sec:conclusion}. 

Throughout this article, we use geometrical units $c=G=M_\odot=1$ unless otherwise stated.

\begin{table*}[t]
  \centering    
  \caption{Configurations. 
    The first column defines the name of the configuration. 
    The subsequent 9 columns describe the corresponding physical properties: 
    EOS, 
    gravitational mass of the individual stars $M^{A,B}$, 
    baryonic mass of the individual stars $M_{b}^{A,B}$, 
    stars' compactnesses $\mathcal{C}^{A,B}$, 
    the total dimensionless quadrupolar tidal coupling constant $\kappa^T_2$,
    the initial dimensionless GW frequency $M \omega_{22}^0$, 
    the ADM mass $M_{\rm ADM}$ and 
    angular momentum $J_{\rm ADM}$ from the initial data. 
    The last two columns describe the resolutions employed and 
    whether the outermost 
    refinement level consists of
    Cartesian boxes or spherical shells, cf.~\cite{Hilditch:2012fp,Dietrich:2016hky}.}
  \begin{tabular}{l||ccccc|cccc||cc}        
  Name & EOS & $M^A$ & $\mathcal{C}^A$ & $M^B$ & $\mathcal{C}^B$ & 
    $\kappa^T_2$ & $M \omega_{22}^0$ &$ M_{\rm ADM}$ &  $J_{\rm ADM}$ & Resolutions & Grid Config.\\
     \hline
     \hline
     MS1b-138138      & MS1b & 1.375008 & 0.1445 & 1.375008 & 0.1445 & 262 & 0.035 & 2.728593 & 8.15792 & (R1,R2,R3) & \texttt{shell} \\
     SLy-138138       & SLy  & 1.375004 & 0.1771 & 1.375004 & 0.1771 & 65  & 0.036 & 2.727632 & 8.00631 & (R1,R2,R3) & \texttt{shell} \\   
     \hline  
     MS1b-153122      & MS1b & 1.527797 & 0.1591 & 1.222228 & 0.1299 & 267 & 0.035 & 2.728700 & 8.03219 & (R1,R2,R3) & \texttt{shell} \\       
     SLy-153122       & SLy  & 1.527388 & 0.1976 & 1.222225 & 0.1573 & 69  & 0.036 & 2.728071 & 7.93386 & (R1,R2,R3) & \texttt{box} \\ 
     \hline
     MS1b-165110      & MS1b & 1.650025 & 0.1708 & 1.100000 & 0.1182 & 282 & 0.035 & 2.729230 & 7.79917 & (R1,R2,R3) & \texttt{shell} \\
     SLy-165110       & SLy  & 1.650019 & 0.2149 & 1.097885 & 0.1416 & 78  & 0.036 & 2.726620 & 7.70012 & (R1,R2,R3) & \texttt{box} \\
     \hline     
     MS1b-150100      & MS1b & 1.500016 & 0.1565 & 1.000001 & 0.1085 & 460 & 0.030 & 2.482511 & 6.65711 & (R1,R2,R3) & \texttt{shell} \\
     MS1b-150100-ecc  & MS1b & 1.500016 & 0.1565 & 1.000001 & 0.1085 & 460 & 0.030 & 2.482230 & 6.61058 & (R1)       & \texttt{box} \\ 
     SLy-150100       & SLy  & 1.500009 & 0.1938 & 1.000007 & 0.1293 & 138 & 0.031 & 2.482086 & 6.58692 & (R1,R2,R3) & \texttt{box} \\
     \hline
     \hline
  \end{tabular}
 \label{tab:config}
\end{table*}

\section{Tidal Effective-One-Body models}
\label{sec:tidalEOB}

The EOB model~\cite{Buonanno:1998gg, Buonanno:2000ef} 
meshes results from the PN approximation, 
valid for any mass ratio but limited to 
weak gravitational fields, 
with knowledge of strong-field effects from the test-particle limit. 
The analytical results are further improved by calibrating 
parameterized higher-order PN terms in the model to NR data to obtain an accurate description of GW signals from 
BBH systems~\cite{Taracchini:2013rva,Pan:2013rra,Damour:2014sva}. 
The conservative EOB dynamics are described by the 
Hamiltonian 
\begin{equation}
H_{\rm EOB}=M\sqrt{1+2\nu\left(H_{\rm eff}-1\right)}-M, 
\end{equation}
with $M=M^A+M^B$ being the total mass and $\nu=M^A M^B/M^2$ the symmetric 
mass ratio. We use $A$ and $B$ to label the two bodies. The effective Hamiltonian $H_{\rm eff}$ describes a particle 
of mass $\mu \equiv \nu M$ moving in an effective spacetime. 
For nonspinning binaries the motion is in a plane and the effective Hamiltonian is given by
\begin{equation}
H_{\rm eff}^2=p_{r*}^2+A\left(1+\frac{p_{\phi}^2}{r^2}+2 \left(4-3\nu \right) 
\frac{p_{r*}^4}{\nu r^2}\right), \label{eq:Heff}
\end{equation}
where $p_\phi$ is the canonical azimuthal angular momentum 
per unit reduced mass, $p_{r*}=p_r/\sqrt{D}$ is a rescaled 
radial momentum, and $A$ and $D$ are potentials describing 
the effective spacetime. The EOB potentials considered in this 
paper20170207 have the form $A=A_{\rm pp}+A_{\rm tidal}$ and $D=D_{\rm pp}$, where 
''${\rm pp}$'' denotes the point-mass part that is used for BBH 
evolutions (\EOBpp). For $A_{\rm pp}$ we use the result given in 
Eqs.~(A1)-(A2h) of Ref.~\cite{Steinhoff:2016rfi} with the 
calibration parameter 
$K=1.712-1.804 \nu-39.77\nu^2+103.166\nu^3$ determined in~\cite{Taracchini:2013rva}. The 
potential $D$ is taken from Eq.~(A4) of 
Ref.~\cite{Steinhoff:2016rfi}. These choices correspond to the model SEOBNRv2\footnote{Note that $A_{\rm pp}$
used for the \EOBre model introduced in Sec.~\ref{sec:tidalEOB:LR}
is slightly different than described above
since the public available code of~\cite{EOB-IHES}
is used, see also~\cite{Barausse:2011dq,Bini:2013rfa,Bini:2013zaa,Bernuzzi:2014owa}.
The effect of the underlying point particle model on the results will be described in
more detail in Appendix~\ref{app:EOBpp}.}. 

We evolve the EOB dynamics using the equations of motion
\begin{subequations}
\begin{equation}
\begin{aligned}
\frac{d r}{dt}&=\frac{A}{\sqrt{D}} \frac{\partial H_{\rm EOB}}{\partial p_{r*}}, \\
\frac{d p_{r*}}{dt}&=-\frac{A}{\sqrt{D}} \frac{\partial H_{\rm EOB}}{\partial r}+{\cal F}_r , \\
\end{aligned}
\end{equation}
\begin{equation}
\frac{d \phi}{dt}=\frac{\partial H_{\rm EOB}}{\partial p_\phi},  \quad
\frac{d p_\phi}{dt}= {\cal F}_\phi. 
\end{equation}
\end{subequations}
The factor $A/\sqrt{D}$ arises 
from the fact that $p_{r*}$ and $r$ are not canonically conjugate variables. 
The gravitational radiation reaction forces 
${\cal F}_\phi$ and ${\cal F}_r$ are constructed from the energy flux through 
\begin{equation}
{\cal F}_\phi=-\frac{1}{\Omega_\phi}\mathcal{\dot E}_{\rm rad}, \ \ \ \ \ {\cal F}_r=\frac{p_{r*}}{p_\phi}{\cal F}_\phi  , \ \ \ \ \ \ \label{eq:EOBforces}
\end{equation}
where the energy flux is computed from 
\begin{equation}
\mathcal{\dot E}_{\rm rad} =\frac{\Omega_\phi^2}{8\pi} \sum_{\ell=2}^8\sum_{m=0}^\ell m^2 |h_{\ell m}^{\rm F}|^2. \label{eq:EOBEdot}
\end{equation}
The sum here is only over positive $m$ since $|h_{\ell -m}^{\rm F}|=|h_{\ell m}^{\rm F}|$. 
The factorized EOB waveforms for point masses are computed from 
\begin{equation}
{h_{\ell m}^{\rm F}}_{\rm pp}=h_{\ell m}^{\rm N}S_{\ell m}T_{\ell m}\rho_{\ell m}^\ell N_{\ell m}.  \label{eq:hlmEOBpp}
\end{equation}
The various factors in Eq.~\eqref{eq:hlmEOB}, are described 
in detail in~\cite{Taracchini:2012ig} and given explicitly in Eqs.~(16), (17), and (A1)-(A14h) therein.

\subsection{Adiabatic tides with 2PN Taylor expanded potential: \EOBad}

Adiabatic tidal effects arise when the distorted NS remains in hydrostatic 
equilibrium. In this case the ratio of its induced multipolar deformation 
to the perturbing tidal field is a constant related to its Love number 
$k_\ell$. The deformation of the NSs changes the energy of the system, which can be captured in the EOB approach by including tidal 
contributions to the EOB potentials of the form
\begin{eqnarray}
A^\text{tidal} &=&-\alpha_2^{(A)} \hat{A}^{(A)}  -\alpha_3^{(A)} \bigg[ 1 + \left(\frac{15}{2}X_A-2\right) u  \nonumber\\
  && + \left(\frac{110}{3}X_A^2-\frac{311}{24}X_A+\frac{8}{3}\right)u^2 \bigg] \nonumber\\
    &&  -\alpha_4^{(A)}+(A \leftrightarrow B). \label{AAT2PN}
\end{eqnarray}
Here $\alpha_2^{(A)} $ is the Newtonian tidal potential given by
\begin{equation}
\alpha_\ell^{(A)}=\frac{2 X_B}{X_A}k_\ell^{(A)}R_A^{2\ell+1}u^{2\ell+2} \label{eq:alpha}
\end{equation} 
where $X_{A,B}=M^{A,B}/M$, $R_A$ is the radius, and $u=1/r$. 
The 2PN corrections to the tidal potential $\hat{A}^{(A)}$ from Eqs.~(6.6) and (6.18) of Ref.~\cite{Bini:2012gu} are given by
\begin{equation}
\hat{A}_{  {\rm adPN}}^{(A)}= 1 + \frac{5}{2}X_A u  
      + \left(\frac{337}{28}X_A^2+\frac{1}{8}X_A+3\right)u^2 \label{AAT}
      \end{equation}
 Tidal effects also influence the dissipative sector since the tidal bulges 
 contribute to the gravitational radiation. In the EOB model this is accounted 
 for by adding to the waveform modes of Eq.~\eqref{eq:hlmEOBpp} a tidal 
 contribution $h_{\ell m}^{\rm tidal}$ to the GW modes so that
 \begin{equation}
 h_{\ell m}^{\rm F}={h_{\ell m}^{\rm F}}_{\rm pp}+h_{\ell m}^{\rm tidal}. \label{eq:hlmEOB}
 \end{equation}
The explicit results for the adiabatic tidal terms $h_{\ell m}^{\rm tidal}$ were worked out 
in Eqs.~(A14)-(A17) of Ref.~\cite{Damour:2012yf} for $l\leq 3$. We use only those terms 
from Ref.~\cite{Damour:2012yf} for which the 
PN knowledge of tidal effects is complete.

\subsection{Dynamical tides with 2PN Taylor expanded potential: \EOBdy}
Dynamical tides become important when the tidal forcing frequency 
due to the companion's orbital motion approaches an eigenfrequency of 
the NS's normal modes of oscillation. The inclusion of the 
effects of dynamic tides from the NS's $f-$modes of oscillation is discussed 
in detail in Ref.~\cite{Steinhoff:2016rfi}. Here, we consider only the 
most practical yet approximate of the models devised therein, where the potential from 
Eqs.~\eqref{AAT2PN} and \eqref{AAT} is used but with $k_\ell$ multiplied by a frequency-dependent 
enhancement factor such that 
$k_\ell \to k_\ell \hat{k}_{\ell \ {\rm dyn}}$ with
\begin{eqnarray}
\hat{k}_{\ell \ {\rm dyn}}&=&a_\ell+b_\ell \bigg[ \frac{\omega_{0\ell}^2 }{\omega_{0\ell}^2-(m\Omega^2)}+\frac{\omega_{0\ell}^2}{2\sqrt{\epsilon_m}\hat t \Omega^\prime (m\Omega)^2}\nonumber\\
&&+\frac{\sqrt{\pi}\omega_{0\ell}^2}{\sqrt{3}\sqrt{\epsilon_m}(m\Omega)^2} {\cal Q}_{\ell m} \bigg] .\ \ \  \label{keffDT}
\end{eqnarray}
Here, $\Omega^\prime=3/8$, $\Omega=M^{1/2}r^{-3/2}$, and $\omega_{0\ell}$ are 
the $f-$mode angular frequencies of the NS, where the subscript $0$ 
indicates that the mode amplitude has no radial nodes.
The quantity ${\cal Q}_{\ell m}$ is
\begin{eqnarray}\label{QlmDT}
{\cal Q}_{\ell m}&=&\cos(\Omega^\prime \hat t^2)\left[1+2\ {\rm F}_{\rm S}\left(\frac{\sqrt{3}}{2\sqrt{\pi}}\hat t \right)\right]\nonumber\\
&&-\sin(\Omega^\prime \hat t^2)\left[1+2\ {\rm F}_{\rm C}\left(\frac{\sqrt{3}}{2\sqrt{\pi}}\hat t \right)\right],
\end{eqnarray}
where the functions ${\rm F}_{\rm S}$ and ${\rm F}_{\rm C}$ are 
Fresnel sine and cosine integrals respectively using the conventions 
in Mathematica. We have defined
\begin{equation}
\hat t=\frac{8}{5\sqrt{\epsilon_m}}\left(1-\frac{r^{5/2}\omega_{0\ell}^{5/3}}{m^{5/3}M^{5/6}}\right),
\end{equation}
and
\begin{equation}
\epsilon_m=\frac{256 \mu  M^{2/3}\omega_{0\ell}^{5/3}}{5m^{5/3}}.
\end{equation}
In Eq.~\eqref{keffDT} a body label $A,B$ on the quantities $\omega_{0\ell}$, $\hat t$, $\epsilon_m$, and 
${\cal Q}_{\ell m}$ is implied. 
For each $\ell$-multipole only $m=\ell$ contributes in Eq.~\eqref{keffDT} because the effect of modes with
 $m<\ell$ has already been taken into account as adiabatic contributions. 
For the lowest order multipoles the coefficients are given by 
$(a_2, a_3,a_4)=(\frac{1}{4},\frac{3}{8},\frac{29}{64})$ and $(b_2,b_3,b_4)=(\frac{3}{4}, \frac{5}{8}, \frac{35}{64})$.

Similar to the treatment for the conservative 
dynamics, the effect of dynamic $f-$mode tides can be incorporated 
in the dissipative sector in an approximate way by multiplying the occurences of 
$k_\ell$ in $h_{\ell m}^{\rm tidal}$ in Eq.~\eqref{eq:hlmEOB} by an effective function 
$k_\ell \to k_\ell \ \hat{k}_{\ell\ {\rm dyn}}^{\rm diss}$. 
For $\ell=2$ this function is given by
\begin{equation}
\hat{k}_{2\ {\rm dyn}}^{(A)\ {\rm diss}}=\frac{\hat{k}_{2\ {\rm dyn}}^{(A)}\left(\omega_{0\ell}^2+6 X_B\Omega^2\right)}{3\Omega^2\left(1+2 X_B\right)}, 
\end{equation}
where $\hat{k}_{2\ {\rm dyn}}^{(A)}$ is the enhancement function 
for body $A$ in the conservative dynamics from Eq.~\eqref{keffDT}. 
We find that similar factors for other modes and from higher multipoles 
have only a small impact on the results and therefore neglect 
them in this study; they are further discussed in Ref.~\cite{TEA}.

\subsection{Adiabatic tides with Gravitational Self-Force-inspired potential: 
            \EOBre}
\label{sec:tidalEOB:LR}

Gravitational self-force calculations have recently computed tidal 
invariants that contain information about strong-field tidal effects in the 
limit of small mass ratios, to linear order in $X_A$ \cite{Dolan:2014pja,Bini:2014zxa}. These results have been augmented in Ref.~\cite{Bini:2014zxa,Bernuzzi:2014owa} by a term 
$\propto X_A^2$ that would describe currently unknown second-order 
self-force effects. Specifically, in this model Eq.~\eqref{AAT2PN} is employed with 
the potential $\hat{A}^{(A)}$ given by
\begin{eqnarray}
\hat{A}_{ {\rm adGSF^+}}^{(A)}&=&1+\frac{3u^2}{\left(1-r_{\rm LR}u\right)} +X_A\frac{a_1^{\rm GSF}(u) }{  \left(1-r_{\rm LR}u\right)^{7/2}}\nonumber\\
&&+X_A^2 \frac{a_2^{\rm 2GSF}(u) }{  \left(1-r_{\rm LR}u\right)^{p}} +(1\leftrightarrow 2)
\label{eq:AGSF}
\end{eqnarray}
The caveats with the choice of gauge for mapping GSF results to EOB potentials 
that contain explicit poles at the light ring $r_{\rm LR}$ are discussed in~\cite{Steinhoff:2016rfi}. 
In Eq.~\eqref{eq:AGSF}, the coefficient $a_1^{\rm GSF}(u) $ is obtained 
from Eqs.~(7.24)--(7.27) of Ref.~\cite{Bini:2014zxa}. 
In the model of Ref.~\cite{Bernuzzi:2014owa} the unknown parameter $a_2^{\rm 2GSF}$ is set to 
$a_2^{\rm 2GSF}(u)=337u^2/28$ and the unknown exponent $p$ is chosen to be $p=4$. The radius of the light ring 
$r_{\rm LR}$ is obtained from the conservative EOB dynamics by solving
\begin{equation}
\tilde{A}(r_{\rm LR})-\frac{r_{\rm LR}}{2}\tilde{A}^\prime(r_{\rm LR})=0, \label{eq:rLRdet}
\end{equation}
with the potential $\tilde{A}=A_{\rm pp}+A^{\rm tidal}_{\rm adPN}$ 
from the PN model in Eqs.~\eqref{AAT2PN} and \eqref{AAT}.

For tests involving the \EOBre model we use the publicly available 
code of~\cite{EOB-IHES} that was also used 
to create the surrogate model in~\cite{Lackey:2016krb}.
Note that this code differs from the EOB model 
described above in the point-particle sector: 
(i) non-quasi-circular (NQC) corrections (the factor $N_{\ell m}$ in the waveform 
modes of Eq.~\eqref{eq:hlmEOBpp} that is necessary to accurately describe the BBH merger) 
are omitted, (ii) a different resummation of the potentials $A_{\rm pp}$ and $D$ 
is used, (iii) the arguments of factors in $h_{\ell m}^{\rm F}$ such as $\rho_{\ell m}$ involve powers of $v_\phi=(\partial H_{\rm EOB}/\partial p_\phi)^{-2/3}\Omega$ evaluated for circular orbits instead of $v=\Omega^{1/3}$, 
(iv) the higher-order logarithms in the tail terms are Pade-resummed, and 
(v) the calibration of the BBH model is different. 
The influence of these differences play only a marginal role compared to the 
tidal effects and how those are incorporated in the EOB model, 
therefore, we will present results obtained with~\cite{EOB-IHES} for \EOBre. 
However, more details about the influence of the underlying point mass models 
will be given in Appendix~\ref{app:EOBpp}.

\section{Binary configurations and numerical methods}
\label{sec:config}

\begin{figure*}[t]
   \includegraphics[width=\textwidth]{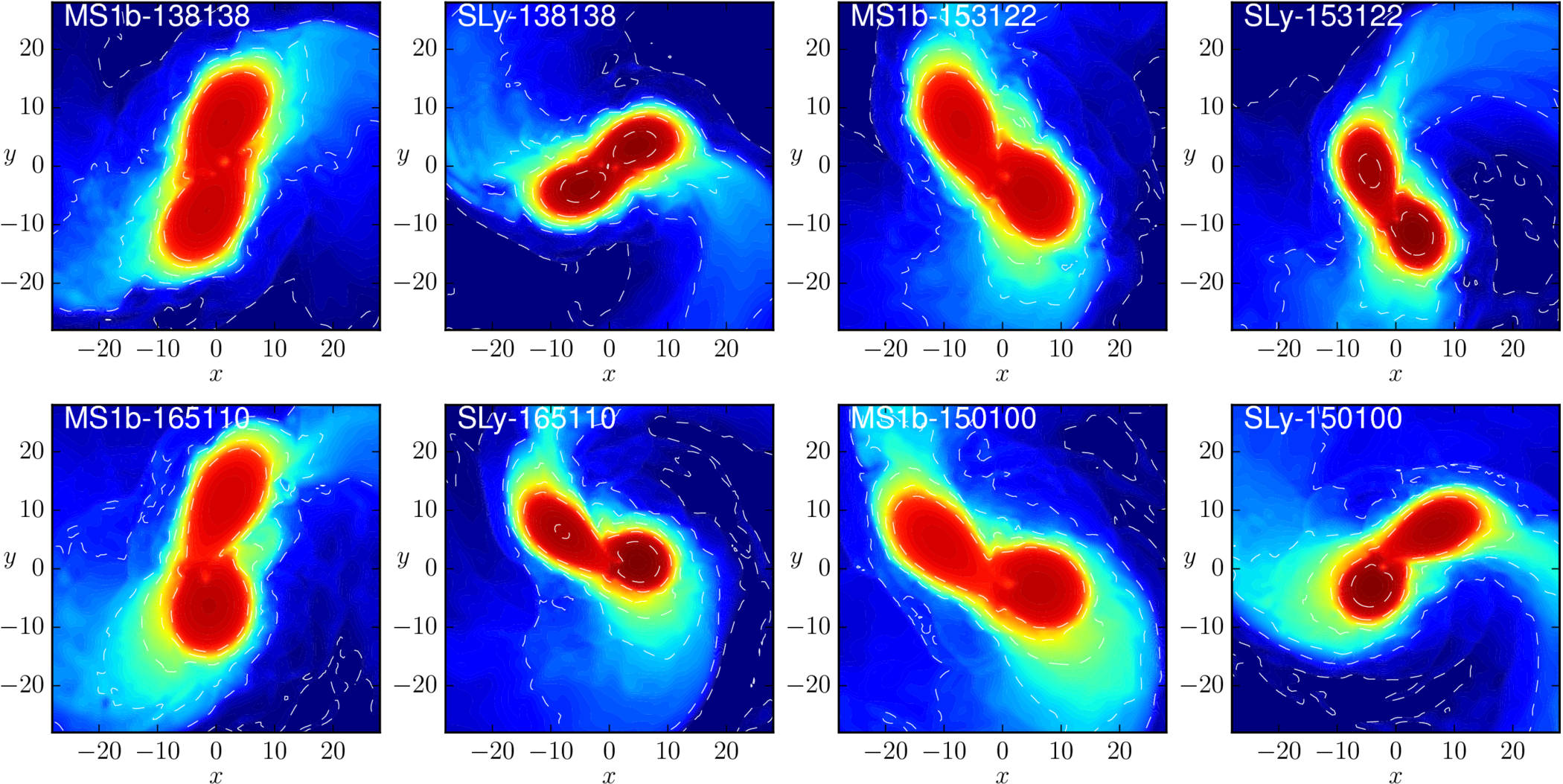}
   \caption{
   Density profile of the setups we investigate (resolution R3). 
   The rest-mass density $\rho$ is shown for all panels 
   on a color bar ranging from $10^{-9} \simeq 6.2 \cdot 10^{8} [{\rm g/cm^3}] $ (dark blue)
   to $10^{-2.8} \simeq 9.8 \cdot 10^{14} [{\rm g/cm^3}]$ (dark red). 
   White dashed lines correspond to values of the density of 
   $10^{-9},10^{-8},10^{-7},10^{-6},10^{-5},10^{-4},10^{-3}$. 
   The $x$-$y$-range is the same in all panels. 
  The plots illustrate that 
   (i) the inner region of the NSs is less dense for stiffer EOSs;
   (ii) the NS radii are larger for stiffer EOSs;
   (iii) the tidal deformation of the lower mass constituent 
   increases with increasing mass ratio;
   (iv) the NSs come into contact before the actual merger (peak in GW amplitude). }
   \label{fig:2d:merger} 
 \end{figure*}  

In this work we consider eight different configurations that have already been studied 
in~\cite{Dietrich:2016hky} with lower resolution. The initial configurations are constructed with
SGRID, see~\cite{Tichy:2009yr,Tichy:2012rp,Dietrich:2015pxa} 
for detailed descriptions.
Dynamical evolutions are performed with the BAM 
code, 
see e.g.~\cite{Brugmann:2008zz,Thierfelder:2011yi,Dietrich:2015iva}. 

We choose two zero-temperature EOSs modeled by 
piecewise polytropes~\cite{Read:2008iy}: 
SLy~\cite{Douchin:2001sv} and MS1b~\cite{Mueller:1996pm}. 
The EOSs are chosen to cover a large range in the BNS parameter space 
with respect to the tidal coupling constant
\begin{equation}
\kappa^T_2 := 2 \left( \frac{q^4}{(1+q)^5} \frac{k_2^A}{C_A^5}  +
\frac{q}{(1+q)^5} \frac{k_2^B}{C_B^5} 
\right)  \  ,
\label{eq:kappa}
\end{equation}
which characterizes Newtonian tidal effects in the EOB Hamiltonian. 
For the dynamical evolution we add an additional  
component to the pressure to describe thermal effects 
$p_{\rm th} = \epsilon \rho (\Gamma_{\rm th}-1)$
with $\Gamma_{\rm th}=1.75$~\cite{Bauswein:2010dn}.
 For both EOSs we focus on four different setups defined by their
total mass 
\begin{equation}
M=M^A + M^B 
\end{equation}
and their mass ratio
\begin{equation}
q = M^A/M^B. 
\end{equation}
The parameters for our setups are 
$(M,q)=$ $(2.75,1.0)$, $(2.75,1.25)$, $(2.75,1.50)$, $(2.50,1.50)$.
To illustrate the effect of changes in the parameters on the merger process we 
show in Fig.~\ref{fig:2d:merger} the 2D density snapshots of these systems at merger, 
which corresponds to the peak in the GW amplitude. The density snapshots 
are gauge-dependent quantities, therefore, Fig.~\ref{fig:2d:merger} 
enables only a qualitative interpretation. 

Building on results from our previous work~\cite{Dietrich:2016hky}, 
we have performed new simulations of the setups at higher resolution. 
Thus, three different resolutions denoted by (R1,R2,R3)
are available for each configuration. The resolutions on the finest grid covering the NSs are 
$(0.23,0.15,0.12)$ for the SLy setups 
and $(0.25,0.16,0.13)$ for MS1b. With these resolutions the 
diameters of the stars are covered by approx.~64, 96, and 128 grid points, respectively. 
Details about the configurations are given in Tab.~\ref{tab:config}. 

To compute results for the periastron advance we consider as a representative 
case the setup MS1b-150100. Reliably extracting this information from NR data, 
as described in Appendix~\ref{app:NRecc}, requires an additional simulation with higher 
eccentricity that we perform following the approach of~\cite{Moldenhauer:2014yaa,Dietrich:2016hky}.

\section{Dynamics}
\label{sec:ej}

\begin{figure*}[t]
   \includegraphics[width=1\textwidth]{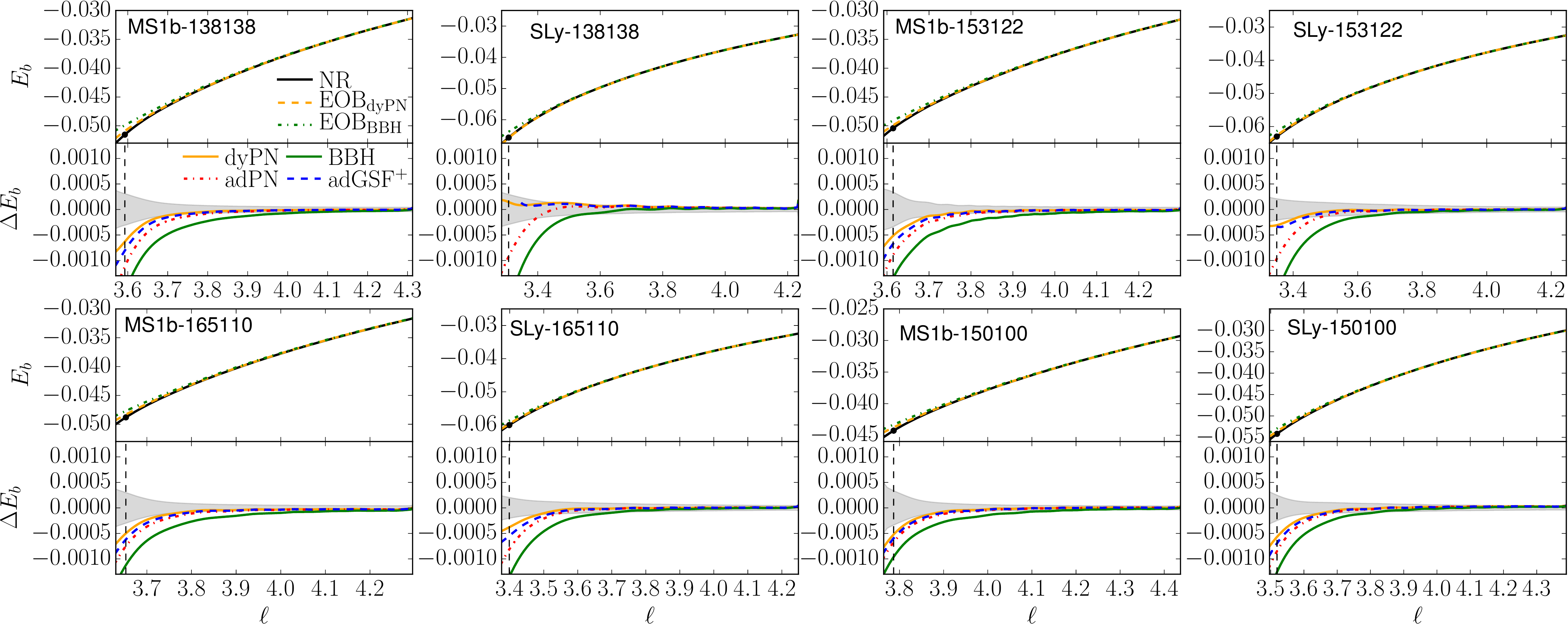}
   \caption{Binding energy $E$ vs.~orbital angular momentum $\ell$ curves. 
            The name of the configuration is shown in the upper left hand corner of the corresponding panel. 
            Upper panels: $E(\ell)$ curves for setups with R3 resolution. 
            We include the EOB estimates 
            \EOBpp (green) and \EOBdy (orange). 
            The moment of merger of the NR results is shown as a circle. 
            Bottom panels: Difference between the EOB results and the NR data. 
            The moment of the NR merger is marked as a vertical black dashed line. 
            The shaded region is the numerical uncertainty estimated from Eq.~\eqref{eq:Eerror}. 
            The ordering of the panels is identical to Fig.~\ref{fig:2d:merger}. 
            }
   \label{fig:EJ} 
 \end{figure*}  

To study the dynamics and energetics of BNS mergers, we present results for the
binding energy vs.~orbital angular momentum of the binary. They were proposed as important 
diagnostic tools to assess the performance of EOB models in~\cite{Damour:2011fu} and were later applied 
for BNS systems in~\cite{Bernuzzi:2013rza}.  
 
\subsection{Binding energy curves in the EOB model}
\label{sec:ej:EOB}

For the binding energy curves we evaluate 
\begin{equation}
E=(H_{\rm EOB}-M)/\mu 
\end{equation}
and 
\begin{equation}
\ell =p_\phi  
\end{equation}
using the solutions to the evolution equations for 
$(r, p_{r*}, p_\phi)$ and obtain the results parametrically. 

\subsection{NR error budget}
\label{sec:ej:NR}

We compute the dimensionless binding energy and angular momentum per
reduced mass as
\begin{eqnarray}
E & = & \left[(M_{\rm ADM}(t=0)- \mathcal{E}_{\rm rad})/M-1\right] \nu^{-1},  \label{eq:Eb} \\
\ell & = & (J_{\rm ADM}(t=0) - \mathcal{J}_{\rm rad})(M^2\nu)^{-1} \ , \label{eq:ell} 
\end{eqnarray}
respectively.
Here, $M_{\rm ADM}$ and $J_{\rm ADM}$ describe the ADM-mass and ADM-angular momentum, and 
$\mathcal{E}_{\rm rad}$ and $\mathcal{J}_{\rm rad}$ are the energy and angular momentum emitted 
from the system via GWs. 
The radiated energy and angular momentum perpendicular to the orbital
plane are calculated as
\begin{align}
\mathcal{E}_{\rm rad}    & = \dfrac{1}{16\pi}\sum_{l,m} \int_{0}^t dt'\left|r\,\dot{h}_{lm}(t')\right|^2 \ , \label{eq:GW_Erad}\\
\mathcal{J}_{z\ \rm rad} & = \dfrac{1}{16\pi}\sum_{l,m} \int_{0}^{t}d t' m\left[r^2\,h_{lm}(t')\dot{h}_{lm}^*(t')\right] \ , \label{eq:GW_Jrad}
\end{align}
with $l_{\rm max}=8$. 

The binding energy vs.~orbital angular momentum curves are directly 
accessible from our numerical simulations, i.e.~no alignment or 
additional shift is needed. This makes 
$E(\ell)$ curves an unambiguous and effective tool 
for a comparisons with analytical models. 

To assess the uncertainty of $E(\ell)$ for our simulations 
we consider four different sources of error: 
\begin{enumerate}[(i)]
 \item \emph{ Uncertainty of the binding energy at $t=0$.} 
      Because of the limited number of points in SGRID and the 
      interpolation of the SGRID initial data on the BAM grid, 
      the initial ADM-mass as well as the individual component 
      masses are only accurate up to $\sim 10^{-5}$.  
      We account for this uncertainty through
      $\Delta E_{\Delta M} = 2 \times 10^{-5}$. 
 \item \emph{Uncertainty of the angular momentum at $t=0$.}
       Similar to the ADM mass, the angular momentum also contains numerical 
       uncertainties of the order of $\Delta J_{\rm ADM} (t) \sim 10^{-3}$. 
       This leads to $\Delta \ell (t=0) = \Delta J_{\rm ADM}(t=0)/(M^A M^B) \sim 10^{-3}$, where 
       we estimate the uncertainty conservatively as $\Delta \ell = 2 \times 10^{-3}$.
       Since $\Delta \ell$ leads to a horizontal shift in the $E(\ell)$ curve we assign a time dependent 
       error on the binding energy of $\Delta E_{\Delta J}(\ell) = E(\ell) - E(\ell-\Delta \ell)$. 
 \item \emph{Finite radius extraction effects.}
       Another source of error is caused by the finite size
       of computational domain and the finite extraction radii.  
       As shown in~\cite{Bernuzzi:2016pie} this error is small ($\lesssim 10^{-1} \rm rad$). 
       To estimate the resulting uncertainty we use two extraction radii, $r_{1}=1000M$ and $r_{2}=750M$, 
       and compute the difference in $E(\ell)$, 
       i.e.~$\Delta E_r(\ell) =  E_{r_{1}}(\ell) - E_{r_{2}}(\ell)$. 
       In future work with more accurate NR data the waveform has to be extrapolated to 
       infinity, but for the scope of this paper20170207 the simple estimate is sufficient. 
 \item \emph{Numerical discretization.} 
       We estimate this error by considering different resolutions. 
       As shown in~\cite{Bernuzzi:2011aq,Bernuzzi:2016pie}
       a conservative error estimate  for the triplet of resolutions employed in this article is 
       to compute the difference between resolution R3 and R2 $\Delta E_{\Delta x}(\ell) = E_{\rm R3}(\ell)- E_{\rm R2}(\ell)$. 
      We find that $\Delta E_{\Delta x}(\ell)$ dominates the overall error in most cases. 
\end{enumerate}

The total uncertainty of the numerical data is approximated by 
\begin{small}
\begin{equation}
 \Delta E_{\rm NR}(\ell ) = \sqrt{\Delta E_{\Delta M}^2 + \Delta E^2_{\Delta J}(\ell) + 
                            \Delta E_{r}^2(\ell) + \Delta E^2_{\Delta x}(\ell)}. \label{eq:Eerror}
\end{equation}
\end{small}

\subsection{Results for $E(\ell)$}
\label{sec:ej:comp}

\subsubsection{General features}

Our results are summarized in Fig.~\ref{fig:EJ}, 
where $E(\ell)$ is shown for the NR data (black), 
the \EOBpp model (green), 
and the \EOBdy model (orange) in the top panels and 
the difference between the NR results and 
the EOB models in the bottom panels. The ordering of the panels is the same as in Fig.~\ref{fig:2d:merger}. 

To aid in the interpretation of our results we first discuss important 
general aspects of the $E(\ell)$ curves. 
Because of the attractive nature of tidal effects BNS systems are more bound than 
BBH systems once systems with the same specific angular momentum are compared. 
Consequently, the binding energy for a given angular momentum 
is smaller for BNS setups than for the corresponding BBH 
case~\footnote{Note that 
for systems with the same frequency 
the opposite statement holds, i.e., the binding energy 
at a given frequency is larger for BBHs than 
for BNSs see the description in~\cite{Dietrich:2016lyp}.}. 

The influence of the EOS on $E(\ell)$ is deduced by contrasting the MS1b and SLy 
curves in adjacent panels of Fig.~\ref{fig:EJ}. We observe that for stiffer EOS (MS1b)
the merger occurs at larger orbital angular momentum and larger binding energy,
so the merger remnant is less bound for MS1b than for SLy.  
This effect is also visible in Fig.~\ref{fig:2d:merger}, where 
MS1b setups merge at larger distances because of their lower compactness.
The influence of the mass ratio is illustrated in Fig.~\ref{fig:EJ} by 
comparing panels with the same EOS (but excluding the 150100 setups that have a lower total mass). 
We see that for increasing mass ratio the moment of merger 
also occurs at higher binding energy and angular momentum. 
Finally, the bottom panels with the same EOS in  Fig.~\ref{fig:EJ} show that an increase in the total mass of the system also leads to a more bound merger remnant. 
Most of these findings can be explained by the quasi-universal properties 
of BNS mergers reported in~\cite{Bernuzzi:2014kca,Bernuzzi:2015eva,Takami:2014tva} and the 
influence of $\kappa_2^T$ on the inspiral and merger dynamics. 

\subsubsection{EOB-NR comparison}

Comparing the EOB and NR curves in Fig.~\ref{fig:EJ} shows that 
for most setups the tidal EOB models deviate from 
the NR predictions shortly before merger. 
This disagreement between NR and EOB is larger 
for stiffer EOSs, smaller total masses, and higher mass ratios, 
i.e.~setups with larger tidal effects. 
Furthermore, it is evident that for all simulations 
\EOBdy and \EOBre stay closer to the NR result than \EOBad. 
This observation is independent of the EOS, mass, and mass ratio. 
The origin of the larger tidal effects for \EOBdy and \EOBre compared to 
\EOBad is different for each model: for 
\EOBdy the dynamical tides enlarge $k_\ell$ when the orbital frequency 
approaches the resonance frequency of the individual stars, while
enhanced tides for \EOBre are caused by approaching the light ring
in the GSF$^+$-potential, Eq.~\eqref{eq:AGSF}. 

Another interesting aspect is the location of the merger point of the individual models:
except for the \EOBre model in the cases SLy-138138 and SLy-153122 (where the \EOBre merger corresponds to the endpoint of the blue dashed curve) the merger for all EOB models occurs at smaller angular momentum and binding energy than in the NR simulations. 
This is interesting since it shows that in most cases the final merger remnant is 
less bound than predicted by the EOB models and that 
additional repulsive effects that are not included in the current EOB 
description become important near merger. 

We emphasize that the NR error quoted here is chosen conservatively, 
i.e., deviations between the NR and the EOB results outside the NR error bars 
mark region where the EOB models must be further improved.  
But note that due to numerical dissipation, which acts as an artificially attractive force, 
the continuum solution will have smaller tidal effects than 
the NR results presented here and will therefore be closer to the EOB models. 
Thus, the correct solution will lie within the lower part of the shaded 
regions in the bottom panels of Fig.~\ref{fig:EJ}.

\section{Periastron Advance}
\label{sec:PA}

\subsection{Periastron Advance in the EOB models}
\label{sec:PA:EOB}

The PA is characterized by the 
ratio of angular and radial frequencies 
$K=\Omega_\phi/\Omega_r$ and, in the case considered here, is influenced by 
a combination of relativistic and tidal effects. In the limit of circular 
orbits, as discussed e.g. in~\cite{LeTiec:2011bk,Hinderer:2013uwa}, the PA for nonspinning 
binaries and for the classes of tidal models considered here can be computed from 
\begin{equation}
K^{-2}=D^{-1}\bigg[A\left(3+\frac{r A^{\prime\prime}}{A^\prime}\right)-2 r A^\prime\bigg],
\end{equation}

\subsection{Extracting the Periastron Advance from NR Simulations}
\label{sec:PA:NR}

\begin{figure}[t]
   \includegraphics[width=0.45\textwidth]{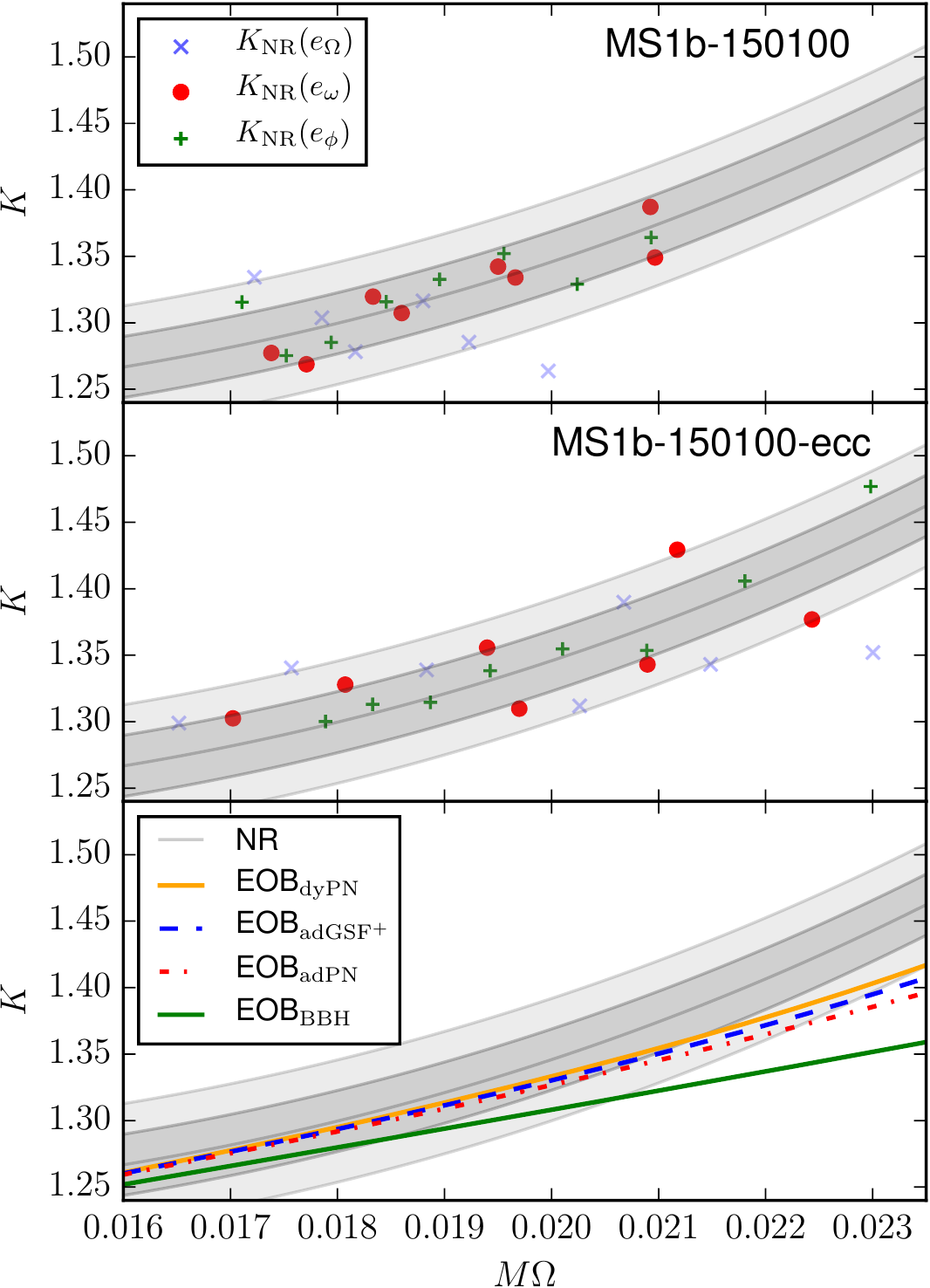}
   \caption{
   The top panel shows the NR results for setup MS1b-150100. 
   Different markers represent results obtained for different eccentricity estimates. 
   The middle panel represents the results for configuration MS1b-150100-ecc. 
   The bottom panel shows the PA for the NR data and all EOB models. 
   The NR data are obtained from a fit to the results $K_{\rm NR}(e_\omega)$, 
   $K_{\rm NR}(e_\phi)$. The shaded intervals mark the $1\sigma$ and $2\sigma$ region of the fit. 
   }
   \label{fig:PA} 
 \end{figure}  

The computation of the PA for NR simulations 
relies on a suitable indicator for the eccentricity.
In this work, we test three different indicators $e_\Omega, e_\omega,e_\phi$ 
based on the 
orbital frequency $\Omega$, the GW frequency of the 
(2,2)-mode $\omega_{22}$, 
and the GW phase $\phi_{22}$ of the (2,2)-mode. 
A detailed description is given in Appendix~\ref{app:NRecc}. 

From the eccentricity indicator we calculate the PA
following the work of~\cite{Mroue:2010re}. 
We determine the times $t_k$ of the extrema of $e$ corresponding to perihelia and aphelia and compute 
the orbital phase $\phi_{\Omega}$ and GW phase $\phi_{22}$ 
at these times $t_k$. Finally the ratio between the angular and radial 
frequency is given by 
\begin{equation}
K =  \frac{\Omega_\phi}{\Omega_r} (t_k) = 
  \frac{\phi_\Omega (t_{k+1}) - \phi_\Omega (t_k)}{2 \pi} = 
  \frac{\phi_{22} (t_{k+1}) - \phi_{22} (t_k)}{4 \pi},
\end{equation}  
where each equal sign corresponds to a different way of extracting $K$ from the NR data. 

\subsection{EOB-NR comparison}
\label{sec:PA:comp}

Extracting $K$ from NR simulations is only possible for sufficiently long 
inspirals. Furthermore, as shown in e.g.~\cite{Mroue:2010re}, 
it is difficult to determine $K$ for 
simulations with almost vanishing eccentricity. 
For these reasons we focus on the 
MS1b-150100 setup which has the lowest starting 
frequency. 
Further we perform a new simulation with higher eccentricity: MS1b-150100-ecc. 
 
Figure~\ref{fig:PA} shows $K$ for all eccentricity measurements for MS1b-150100
in the upper panel, and the results for MS1b-150100-ecc in the middle 
panel\footnote{We have simulated MS1b-150100-ecc only for resolution R1. 
The resolution has only a small influence on the measurement of the PA, since 
resolution effects become dominant at later times of the simulation, 
where no eccentricity can be measured at all.}.
We find that the scattering of data points obtained from $K_{\rm NR}(e_\phi)$ 
is smallest and that these results are in agreement with $K_{\rm NR}(e_\omega)$.
By contrast the results from $K_{\rm NR}(e_\Omega)$ are more scattered and 
do not help to put constraints on $K$.
Thus, we combine all data points obtained with $K_{\rm NR}(e_\phi)$ and 
$K_{\rm NR}(e_\omega)$ and fit the NR data for both configurations according to 
\begin{equation}
K_{\rm NR}(M\Omega)  = \frac{c_0 +c_1 (M \Omega) + c_2 (M \Omega)^2 } 
{ \sqrt{1- 6 (M \Omega )^{\frac{2}{3}}}},
\end{equation}
cf.~\cite{Hinderer:2013uwa}. 
We indicate the $1\sigma$ and $2\sigma$ region of the fit 
as dark and light shaded regions respectively in Fig.~\ref{fig:PA}. 

The bottom panel of Fig.~\ref{fig:PA} illustrates that 
the NR data contains larger tidal effects than those predicted by the EOB models. 
Because of the difficulties in reliably computing the NR error for the PA 
(as discussed in Appendix~\ref{app:NRecc}, see also~\cite{Hinderer:2013uwa} for 
uncertainty estimates of PA in BBH setups) the results presented here have to be considered with caution. 
However, it is clearly visible from Fig.~\ref{fig:PA} that for low frequencies $M\Omega<0.019$ 
the PA predicted from the EOB models and the results from the NR simulations agree very well, 
cf.~dark shaded region. 
At higher frequencies the curves start 
to deviate and the point-mass result \EOBpp becomes distinguishable from the NR uncertainty region. The 
\EOBpp result leaves the $1\sigma$ region at $M \Omega \approx 0.019$ 
and the $2\sigma$ region at $M\Omega \approx 0.021$. 
This demonstrates for the first time that current NR simulations are able 
to resolve tidal effects on the PA. 
Up to frequencies of order $M\Omega \approx 0.023$ the results from the EOB models \EOBdy 
and \EOBre remain within the $2\sigma$ region of 
the NR data. Beyond this frequency no additional NR data points 
(see upper and middle panel) are available 
and, thus, no further statement to discriminate between different tidal EOB models can be made.

\section{Waveforms}
\label{sec:GWs}

\subsection{EOB waveforms}
\label{sec:GWs:EOB}

The EOB waveforms are computed by evaluating the functions $h_{\ell m}^{\rm F}$ in Eq.~\eqref{eq:hlmEOB}. 
They depend on the EOB trajectory and enter into the backreaction of GW losses 
on the dynamics through Eq.~\eqref{eq:EOBforces}. To compute the time evolution of 
the modes we therefore solve the algebraic 
equations~\eqref{eq:hlmEOB} jointly with the EOB equations of motion.

\subsection{NR error budget}
\label{sec:GWs:NR}

Gravitational waves are extracted using the Newman-Penrose formalism~\cite{Newman:1962} 
from the curvature scalar $\Psi_4$. Individual modes can be obtained by a multipole decomposition on
extraction spheres using spherical harmonics with spin weight $-2$. 
The algorithm and implementation is described in detail in~\cite{Brugmann:2008zz}.
We reconstruct the metric multipoles $r h_{l m}$ from the curvature multipoles 
$r \Psi_{4\ l m} $ using the frequency domain integration
of~\cite{Reisswig:2010di} with a cutting frequency of $m \omega_{22}^0/2$, 
see Tab.~\ref{tab:config}.

\subsubsection{$\phi(t)$-Evolution}

The NR waveforms are presented as a function of the retarded time 
\begin{equation} 
 u=t-r_*=t-r_{\rm extr}-2M\ln\left[r_{\rm extr}/(2M)-1\right]. \label{eq:uretarded}
\end{equation}
For the extraction radius we use the value $r_{\rm extr}=1000$. 
As shown in~\cite{Bernuzzi:2016pie} for this radius the error due to using 
waveforms extracted at a finite radius is of order $\sim 0.1$rad and decreases 
during the simulation because the GW frequency increases. 
As a simple error estimate we compute 
$\Delta \phi_{r} = \phi_{r_1} - \phi_{r_2}$ with $r_1=1000$ and $r_2=750$,
similar to the investigation of the binding energy discussed above.

In most simulations the dominant source of uncertainty are truncation errors during 
the last few orbits before merger. 
We estimate the uncertainty due to finite resolution 
by the difference between setups R3 and R2, 
$\Delta \phi_{\Delta x } = \phi_{\rm R3}-\phi_{\rm R2}$. 
As shown in~\cite{Bernuzzi:2011aq,Bernuzzi:2016pie} for these resolutions this
is a conservative measure of the error. 

Finally, we also take into account that mass is not exactly conserved in the simulations.  
In particular, as discussed in~\cite{Dietrich:2015iva}, the treatment of the atmosphere 
can cause mass loss or gain 
and is a source of systematic errors. 
The dephasing caused by variations in the mass 
is taken into account as 
$\Delta \phi_{\Delta M} = \omega u\Delta M_b /M_b$, 
where $M_b$ denotes the baryonic mass of the system at the beginning of the simulation 
and $\Delta M_b$ is the change in the baryonic mass. 

The total phase uncertainty is then given by 
\begin{equation}
 \Delta \phi = {\rm{max}}_{\tilde{u}\in [0,u]}\sqrt{\Delta \phi_{r}^2 (\tilde{u}) + 
                     \Delta \phi_{\Delta x}^2 (\tilde{u}) +  
                     \Delta \phi_{\Delta M}^2 (\tilde{u}) }.\label{eq:nrerror:phit}
\end{equation}

We take into account that some errors are non-monotonic by using  
${\rm{max}}_{\tilde{u}\in [0,u]}[...]$. 
This ensures that the total error is always increasing over 
time~\footnote{Note that we do not include the influence of the residual 
eccentricity in our error budget since this would require the use of eccentricity reduced waveforms. 
But as presented in~\cite{Dietrich:2015pxa} the remaining eccentricity can 
cause phase differences of the order of $\sim 0.2$rad.}. 

\subsubsection{$\phi(\omega)$- Evolution}

In addition to the time evolution of the GW phase, we also present
the phase accumulated over a given frequency interval. 
For all setups we compute the accumulated phase 
with respect to the reference frequency $M\omega_{22}=0.04$, which corresponds to 
$470 \rm Hz$ for $M=2.75$ and to $517 \rm Hz$ for $M=2.50$. The 
details of the computation are discussed in Appendix~\ref{app:phiomega}.

The error budget of $\phi(\omega)$ contains the following components, similar to 
the analysis of the time evolution $\phi(t)$: 
\begin{enumerate}[(i)]
 \item  We take into account the fact that the GWs are extracted at finite radii by considering
       $\Delta \phi_r (\omega) = \phi_{r1} (\omega) - \phi_{r2}(\omega)$ with $r_1=1000$ and $r_2=750$.
 \item The uncertainty caused by numerical discretization is estimated as 
       $\Delta \phi_{\Delta x } (\omega)= \phi_{\rm R3} (\omega) -\phi_{\rm R2}(\omega)$.
 \item We include an estimate resulting from the non-conservation of the total mass
       $\Delta \phi_{\Delta M} (\omega)$. 
 \item An additional source of error is the uncertainty caused by the low-pass filtering of the raw NR data 
       (Appendix~\ref{app:phiomega}). We account for this by varying the cutoff-frequency 
       by a factor of $2$ and obtain the estimate
       $\Delta \phi_{\Delta \omega_{\rm cut}}(\omega) = 
         \phi_{\omega_{\rm cut}} (\omega)- \phi_{2 \cdot \omega_{\rm cut}}(\omega)$.
 \item A further contribution to the error budget arises from the fact that the 
       phase evolution is computed with respect to a phase at $M \tilde{\omega}_{22}= 0.04$. 
       The uncertainty in this reference value is estimated to be
       $\Delta \phi_{M \tilde{\omega}} (\omega)= \phi_{M \tilde{\omega}}(\omega) - 
       \phi_{M \tilde{\omega}\pm2.5\times 10^{-4}}(\omega)$. 
\end{enumerate}

The total error is given as: 
\begin{align}
 \Delta \phi(\omega) = & \left[ \Delta \phi_r^2 (\omega) + 
                              \Delta \phi_{\Delta x }^2 (\omega) + 
                              \Delta \phi_{\Delta M}^2 (\omega)
                      \right. \nonumber\\ 
                      & \left. +  \Delta \phi_{\Delta \omega_{\rm cut}}^2(\omega) +  
                               \Delta \phi_{M \tilde{\omega}}^2 (\omega) 
                      \right]^{1/2} \label{eq:nrerror:phiomega} . 
\end{align}

\subsection{EOB-NR comparison: $\phi(t)$}
\label{sec:GWs:phi(t)}

\begin{figure*}[t]
   \includegraphics[width=\textwidth]{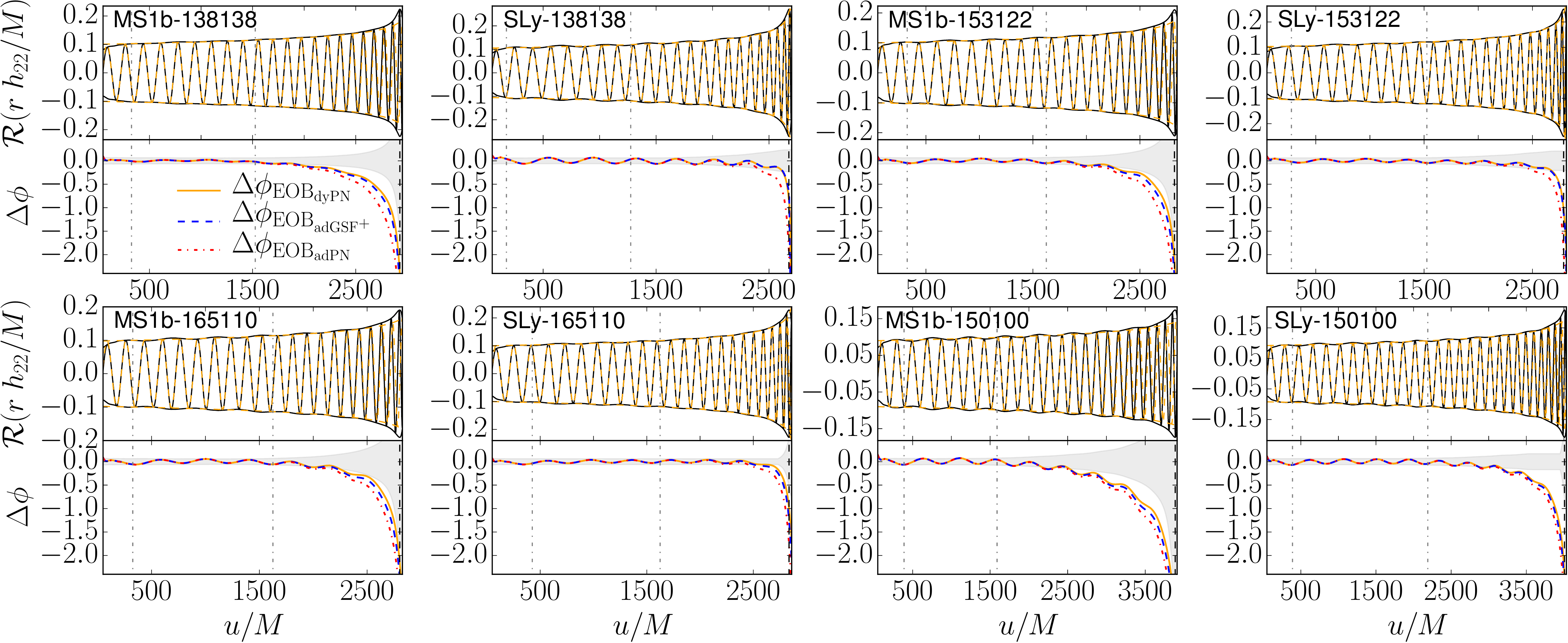}
   \caption{Dominant (2,2)-mode of the GW for all configurations. 
            The top panels represent the real part and the amplitude of the NR data (black)
            and the \EOBdy model (orange). 
            The vertical dash-dotted lines show the interval 
            over which the EOB and NR waveform are aligned. 
            The bottom panels represent the dephasing 
            $\Delta \phi = \phi_{\rm EOB} - \phi_{\rm NR}$ in rads for the tidal EOB models. 
            The shaded region represents the estimate of the uncertainty in the NR results, 
            and the vertical dashed line marks the moment of merger from NR. 
            Notice that the ordering of the panels is identical to Fig.~\ref{fig:2d:merger}.}
   \label{fig:GW22} 
 \end{figure*}  

As shown in e.g.~\cite{Dietrich:2016hky} more than $99\%$ 
of the total energy emitted in a BNS merger up to a mass ratio of $q=1.5$
is contained in the (2,2) mode. 
This motivated previous EOB and NR comparisons 
to focus solely on this mode. 
However, we want to go beyond previous work and include the 
(2,1), (3,3), and (4,4) mode in our comparison. 

By contrast to the comparisons of results for $E(\ell)$, $K$, and $\phi(\omega)$ discussed below, we 
do not include the \EOBpp model in our analysis of $\phi(t)$. 
The reason is that a robust comparison of the dephasing over time requires aligning two waveforms 
over a time interval where they agree. 
However, our simulations cover only the last $\sim 10$ orbits 
before merger, where tidal effects \textit{cannot} be neglected and a 
reliable alignment of BBH and BNS waveforms cannot be achieved. 

\subsubsection{Dominant (2,2)-mode}
 
Figure~\ref{fig:GW22} presents the results for the time evolution of the 
dominant (2,2)-mode. In the upper panels the real part of 
$r h_{22}$ is shown for the NR data (black) and 
the \EOBdy model (orange dashed), we also include the amplitude $|rh_{22}|$ as thin lines. 
The time interval used for the alignment is marked by vertical gray dash-dotted 
lines\footnote{For large $\kappa_2^T$ 
changing the alignment interval effects $\Delta \phi(t)$ and 
a more robust comparison would require longer NR simulations.}. 
In the bottom panel we present the difference between the NR waveform 
and the tidal EOB models. 
The moment of merger in the NR simulation is shown as a vertical black 
dashed line in the bottom panels. 
The numerical uncertainty is shown as a shaded region. 

We conclude from Fig.~\ref{fig:GW22} that:
\begin{enumerate}[(i)]
 \item  For setups with a small tidal coupling constant 
        the dephasing between EOB and NR waveforms 
        is small and remains within the 
        numerical uncertainty almost up to the merger.  
 \item  Variations in the mass ratio have a negligible effect on the performance of the 
        EOB models when considering $\phi(t)$. 
 \item  For setups with stiffer EOSs the 
        EOB models start to deviate from the 
        NR simulation a few orbits before the merger.
 \item  In all cases the EOB models underpredict tidal effects, 
        i.e.~$\phi_{\rm EOB}(u) < \phi_{\rm NR}(u)$.
 \item  The EOB merger generally occurs after the NR merger. 
 \item  For all setups the performance of the tidal EOB models is similar, with 
        the \EOBdy and \EOBre models remaining closer to the NR results than the \EOBad model.
\end{enumerate}

\subsubsection{Higher modes} 
 
 \begin{figure*}[t]
   \includegraphics[width=\textwidth]{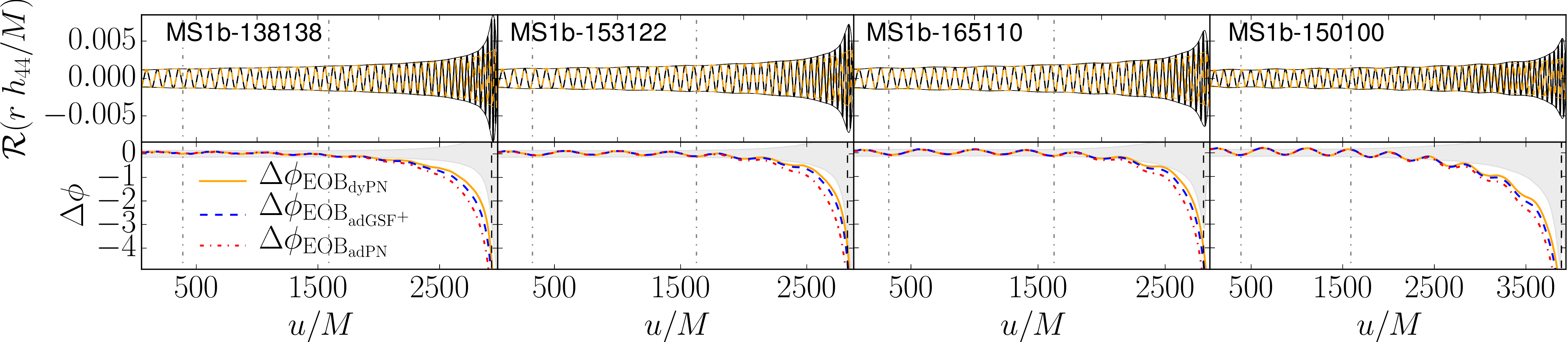}
   \caption{(4,4)-mode of the GW for MS1b configurations. 
            The top panels represent the real part and the amplitude of the NR data (black)
            and the \EOBdy model (orange). 
            The vertical dash-dotted lines show the interval 
            used to align the EOB and NR waveform. 
            The bottom panels represent the dephasing 
            $\Delta \phi = \phi_{\rm EOB} - \phi_{\rm NR}$ in rads for the tidal EOB models. 
            The shaded region represents the estimate of the uncertainty of the NR results. 
            The vertical dashed line marks the moment of merger for the NR result.}
   \label{fig:GW44} 
 \end{figure*}   
 
 \begin{figure*}[t]
   \includegraphics[width=\textwidth]{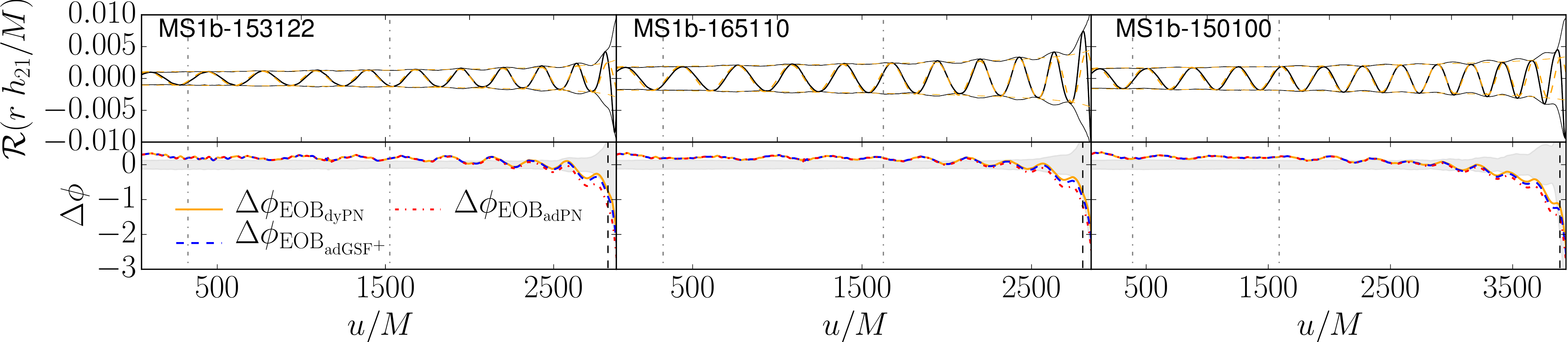}
   \includegraphics[width=\textwidth]{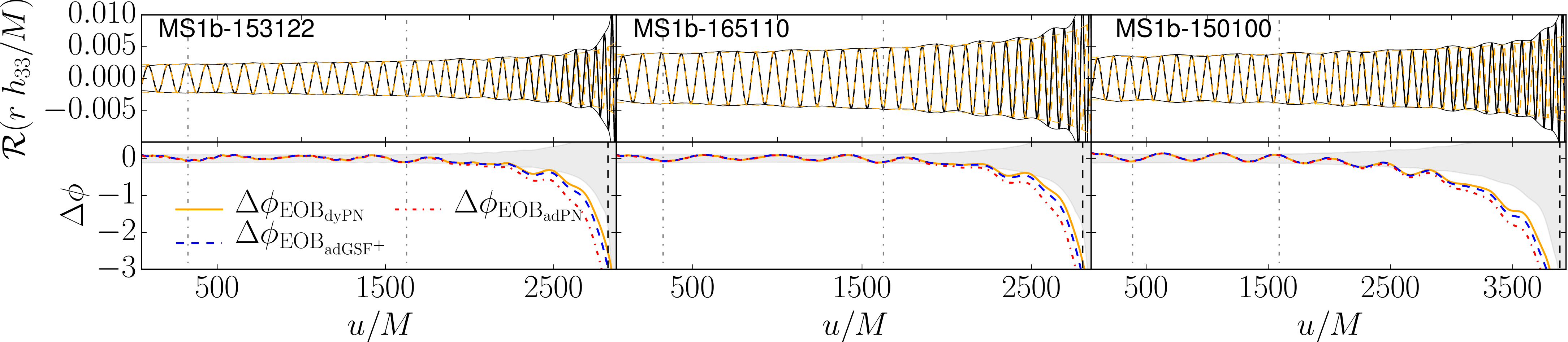}
   \caption{(2,1)-mode (top panels) and (3,3)-mode (bottom panels) of the GW signal for 
            the unequal mass setups employing the MS1b EOS. 
            The top panels represent the real part and the amplitude of the NR data (black)
            and the \EOBdy model (orange). 
            The vertical dash-dotted lines show the interval 
            used to align the EOB and NR waveform. 
            The bottom panels represent the dephasing 
            $\Delta \phi = \phi_{\rm EOB} - \phi_{\rm NR}$ for the different tidal EOB models in rads. 
            The shaded region represents the estimate of the uncertainty of the NR results. 
            The vertical dashed line marks the moment of merger for the NR result.}
   \label{fig:GW2133} 
 \end{figure*} 

While for equal mass configurations the (2,1) 
and (3,3) modes are zero, they are non-zero for unequal masses. 
The energy emitted in the (2,1) and (3,3) mode increases almost 
linearly for an increasing mass ratio. 
By contrast the amplitude and consequently the emitted 
energy of the (4,4) mode is almost independent of the 
mass ratio, see~\cite{Dietrich:2016hky} for further details. 
We present the (4,4) mode for all models employing the 
MS1b EOS in Fig.~\ref{fig:GW44}\footnote{Note that we do not study
systems employing the SLy EOS since those have (as presented in~\cite{Dietrich:2016hky}) a 
larger drift of the center of mass, which effects the computation of the 
subdominant modes.}. 
The (2,1) and (3,3) mode for the unequal 
mass setups with MS1b EOS are shown in 
Fig.~\ref{fig:GW2133}. We used the alignment based on the $(2,2)$ mode, where we obtained the time and phase shifts $(\delta t^{\rm align}, \delta \phi_{22}^{\rm align})$. This determines the shifts for the other $(\ell, m)$ modes to be $\delta\phi_{\ell m}^{\rm align}=m\delta \phi_{22}^{\rm align})/2$, together with the same time shift. 
Another consideration is that the point-mass EOB model for higher modes has not been calibrated to NR results, e.g. there is no equivalent function to $N_{22}$. As shown in  Appendix~\ref{app:EOBpp}, the influence of such calibrations of the EOB model is $\lesssim 0.03$rad for the MS1b setups and thus much smaller than the size of tidal effects. 

Overall we find that the dephasing for the higher modes 
is similar to the results obtained for the dominant (2,2) mode. 
The performance of the EOB models depends only weakly on the mass ratio
since $q$ has only a small influence on the tidal coupling constant. 
By contrast, the total mass impacts the accuracy of the EOB models since 
tidal effects strongly decrease with increasing total mass. For example, for MS1b-150100 where $\kappa_2^T$ 
is $\sim 180$ larger than for MS1b-165110 
the dephasing from NR is larger since the larger tidal effects are inadequately captured by the models.

Considering the agreement of the amplitude between the EOB models and the NR results, 
we find that close to the moment of merger the amplitude predicted 
by the EOB model is smaller than those for the NR simulation. 
This is true for all setups and is a robust (qualitative) observation. 
However, since the computation of the subdominant modes is challenging and 
the numerical simulations are more inaccurate close to the 
merger a more quantitative investigation 
is postponed to future work, where simulations with higher resolution and 
waveforms corrected for the center of mass drift will be available.
 
\subsection{EOB-NR comparison: $\phi(\omega)$}
\label{sec:GWs:phi(omega)} 
 
\begin{figure*}[t]
   \includegraphics[width=\textwidth]{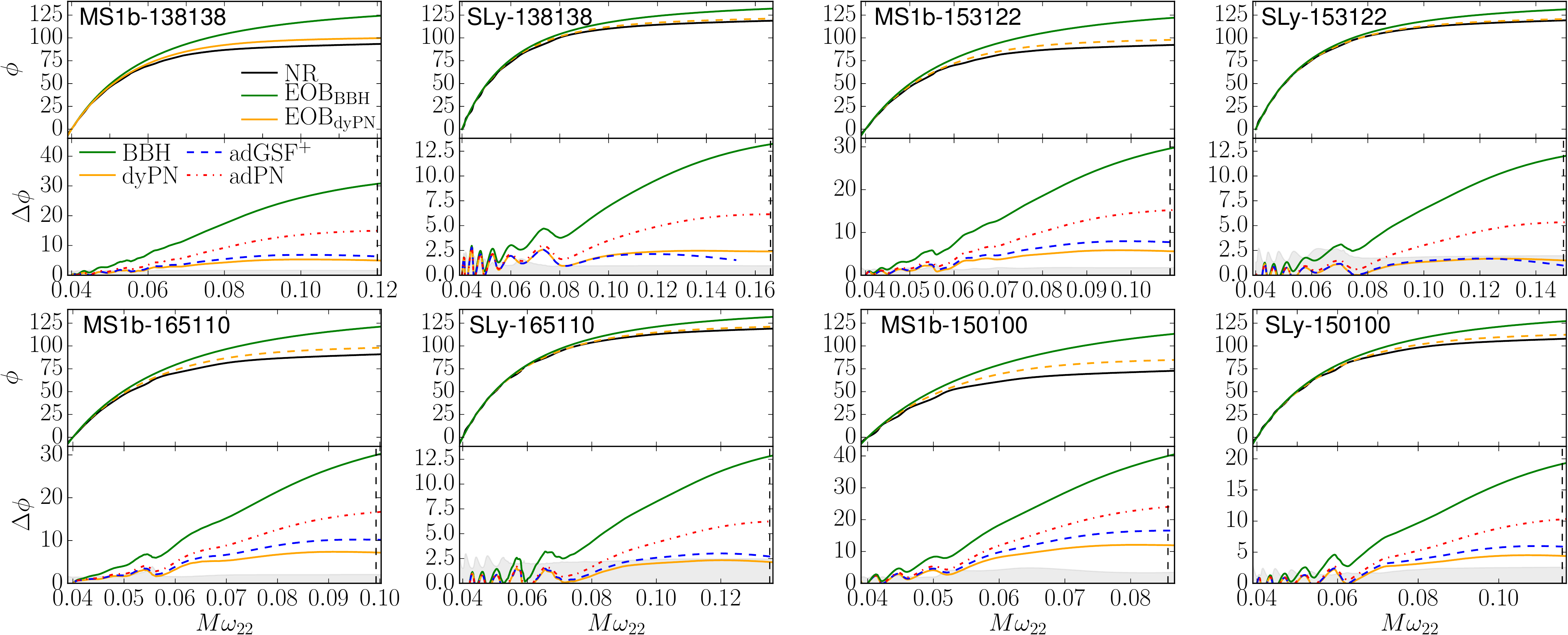}
   \caption{Phase evolution as a function of GW frequency. 
   The top panels show the evolution of the phase of the (2,2) mode 
   for the NR data (black), the \EOBpp model (green), 
   and the \EOBdy model (orange) as a function of the dimensionless GW frequency $M \omega_{22}$. 
   The bottom panels show the difference $\Delta \phi = \phi_{\rm EOB} - \phi_{\rm NR}$ 
   for all EOB models in rads. 
   The shaded region represents the estimate of the uncertainty of the NR results. 
   The vertical dashed lines in the bottom panels mark 
   the moments of merger for the NR data. }
   \label{fig:phiomega} 
 \end{figure*}  
 
Another important means to assess the performance of EOB models 
is the phase evolution as a function of the frequency. 
The results of this comparison are shown in Fig.~\ref{fig:phiomega}. The top panels 
exhibit the accumulated phase $\phi(\omega)$ with respect to the reference frequency $M \omega_{22}=0.04$ 
computed for the \EOBpp model (green), the \EOBdy model (orange), 
and the NR data (black). The difference between the NR and the EOB results is shown in the bottom panels. 
In contrast to the small tidal contributions in the time evolution of the phase $\phi(t)$, 
the accumulated phase as a function of frequency due to tidal effects is of the order of $10$ to $40$rad 
for our setups, which enables an easier assessment of tidal effects. 
Furthermore, because no alignment is needed for comparisons of $\phi(\omega)$, 
the point-mass results \EOBpp can readily be included in the comparison. 

The qualitative behavior of $\phi(\omega)$-plots is the following. 
Because the frequency increases during the inspiral, the accumulated phase within 
a given frequency interval $\Delta (M\omega)$ decreases, i.e.~$\partial_\omega^2 \phi(\omega)<0$.
The effect of the EOSs (seen in Fig.~\ref{fig:phiomega} by comparing SLy and MS1b configurations) 
is that MS1b setups merge at lower frequency due to their smaller compactness. 
This is in agreement with the conclusions drawn from the results of the
binding energy in Sec.~\ref{sec:ej}. 
Considering the difference $\phi_{\rm NR} - \phi_{\rm EOB_{\rm BBH}}$, shown as the 
green curves in the lower panels of Fig.~\ref{fig:phiomega}, we see that smaller total masses 
lead to a larger accumulated phase due to the larger tidal effects.  The oscillatory behavior 
of $\Delta \phi$ for low frequencies is caused by the residual eccentricity that is
not entirely eliminated by the low-pass filter, cf.~Appendix~\ref{app:phiomega}. 

In contrast to the results for $\phi(t)$ shown in Fig.~\ref{fig:GW22} the results for 
$\phi(\omega)$ in Fig.~\ref{fig:phiomega} clearly exhibit a substantial 
improvement of the \EOBdy and \EOBre models over the \EOBad model that 
has a larger discrepancy to the NR results.
As described before the agreement between EOB and NR is generally better 
for softer EOSs for which tidal effects are smaller. 
But when considering the evolution of $\phi(\omega)$, 
where differences are more pronounced, it is apparent that several 
models are outside the NR uncertainty. 
Only for the cases SLy-153122 and SLy-165110 
are the \EOBdy results in agreement with the NR waveforms to within the current uncertainty in NR data. 
Interestingly, we observe that for some setups the dephasing of the \EOBre 
model decreases close to the merger, and, in one case terminates 
before attaining the merger frequency. 
These features indicate that the \EOBre model likely overestimates 
tidal effects in the late inspiral for these particular cases. 
 
\section{Current problems and future prospects} 
\label{sec:case}

\subsection{Test cases}

Having presented the results of comparing EOB predictions to NR 
results for a variety of systems in the previous sections, 
we next perform a more detailed analysis to identify the source of the discrepancies. 
For simplicity, we focus on two complementary cases at opposite 
corners in the parameter space considered here: 
SLy-138138, where both stars have a soft EOS and equal masses,
and MS1b-165110, where both stars have a stiff EOS and the mass ratio is $q=1.5$.

\subsubsection{SLy-138138}

 \begin{figure*}[t]
   \includegraphics[width=0.95\textwidth]{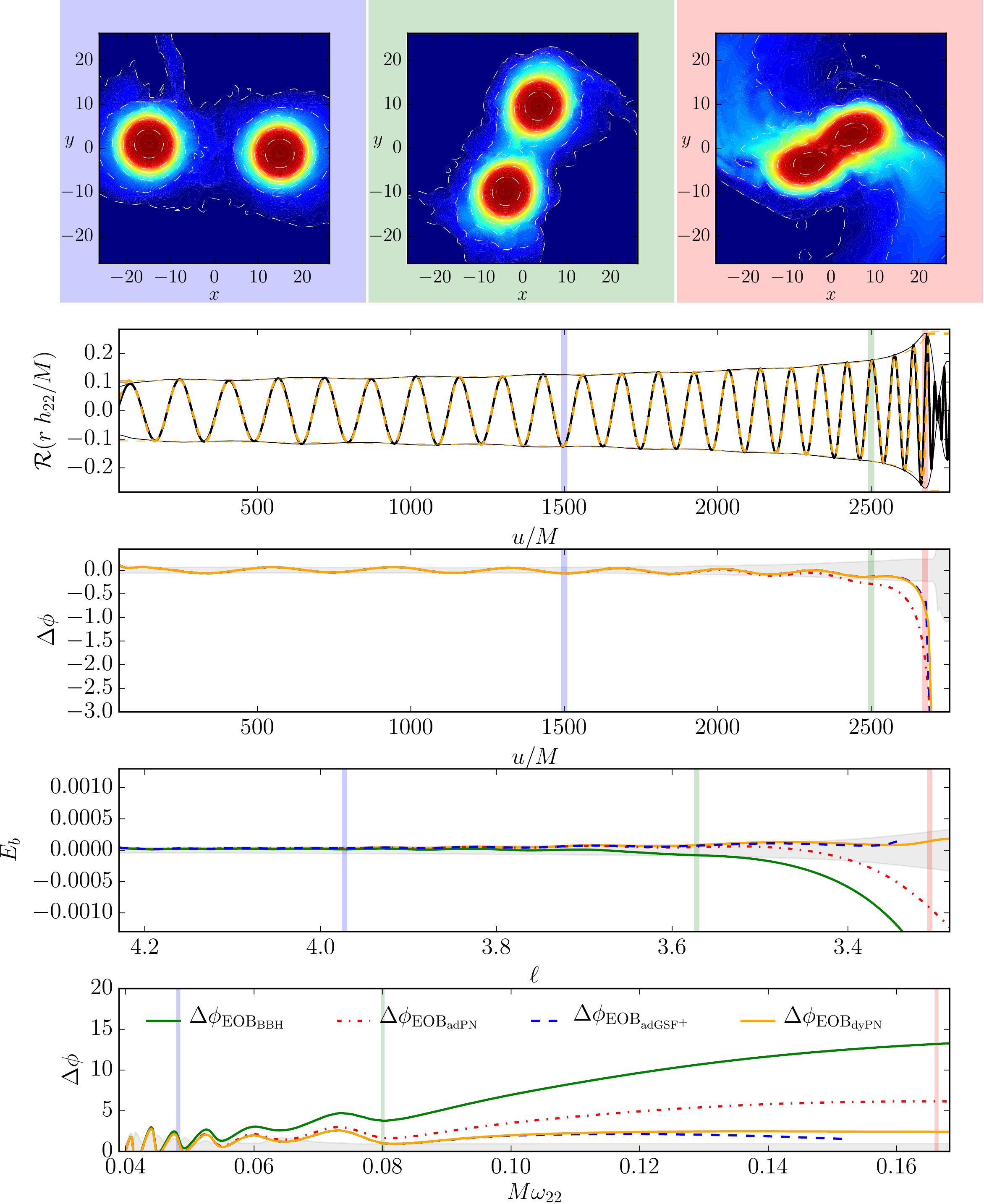}
   \caption{The top row shows rest-mass density snapshots with 
            the color bar ranging from $10^{-9}$ (dark blue)
            to $10^{-2.8}$ (dark red) for SLy-138138. 
            White dashed lines show contour densities lines with
            $10^{-9},10^{-8},10^{-7},10^{-6},10^{-5},10^{-4},10^{-3}$. 
            The snapshots correspond to times $t=1500M$ (blue),
            $2500M$ (green), and the moment of merger $2675M$ (red).
            The second row shows the real part and the amplitude of the 
            GW for the NR data (black) and \EOBdy (orange). 
            The third row shows the the dephasing of the EOB models with 
            respect to the NR simulation. 
            The fourth row shows the difference of the binding energy 
            between the EOB models (including green \EOBpp) and the NR simulation. 
            The fifth row shows the dephasing as a function of the frequency.
            In all panels we use the same color scheme for all models. `
            The times corresponding to the snapshots are marked as shaded regions. 
            The $E(\ell)$ and $\phi(\omega)$ curves end at the moment 
            of merger for all EOB models, which is the reason why for \EOBre 
            $E(\ell)$ and $\phi(\omega)$ terminate before the NR merger.}
   \label{fig:case:SLy-138138} 
 \end{figure*}

In Fig.~\ref{fig:case:SLy-138138} we summarize important quantities for 
SLy-138138. In the top panels we show three snapshots of the rest-mass density $\rho$
obtained from the NR simulation, where the color bar is identical 
to that used in Fig.~\ref{fig:2d:merger}. 
The snapshots correspond to times $t=1500M$ (blue),
$2500M$ (green) and the moment of merger $2675M$ (red).
The other panels show the real part and amplitude of the GW, 
the dephasing of the EOB models with respect to the NR simulation, 
the difference of the binding energy between the EOB models and 
the NR simulation, and the dephasing as a function of the frequency. 
The retarded time, angular momentum, and frequency corresponding the density snapshots are marked as 
columns of the same color in the plots. 

For setup SLy-138138 both stars have a 
soft EOS and the same gravitational mass of $M=1.375$. 
The tidal coupling constant at $t=0$ is $\kappa_2^T=65$. 
Within our data set this configuration has the smallest tidal effects. 
Consequently, differences between the EOB models 
and the NR simulation are also small. 

At time $t=1500M$ (left density snapshot, blue columns) the two stars 
are well-separated and no tidal deformation is visible. 
The low density regions ($\rho \lesssim 10^{-8}$, dark shades of blue) 
in the plot are affected by the artificial
atmosphere in the NR treatment and no clear star surface exists. 
The imprint of the artificial atmosphere decreases for increasing resolution. 
At this time the tidal EOB models are in agreement with the NR 
simulation for all quantities considered. 

As the system evolves and the separation decreases tidal effects 
become larger. At $t=2500M$ (middle density snapshot, green columns) 
the tidal deformation of the stars is visible.
At this stage the GW phasing predicted by the \EOBad model starts to deviate from the NR 
data as seen in the plots of $\Delta \phi(t)$ and $\Delta \phi(\omega)$. 
This indicates that the \EOBad model underestimates tidal effects. 

Approximately $1.5$ orbits later the GW amplitude reaches its maximum 
and the stars merge (red frame/columns). Because of the enhanced tidal effects in 
\EOBre and \EOBdy these models are able to describe the GW phase as a function of time up 
to the last GW cycle before the merger. 
We note three important observations:
(i) The model \EOBre predicts a merger at a higher specific orbital angular momentum 
(endpoint of the blue dashed curve in the fourth panel) than the model \EOBdy. 
The merger of the NR simulation lies between both models. 
(ii) The dephasing $\Delta \phi_{\rm EOB_{\rm adGSF^+}}(\omega)$ decreases 
shortly before merger, which is caused by the fact that 
the binary separation approaches the light ring and tidal effects are overestimated at this stage.
(iii) Since tidal effects are overall small, both models \EOBre and
\EOBdy give a reasonably accurate representation of the dynamics and waveform. 
Further improvements of tidal EOB models in this part of the parameter 
space would require more accurate NR simulations. 

\subsubsection{MS1b-165110}
 
\begin{figure*}[t]
   \includegraphics[width=0.95\textwidth]{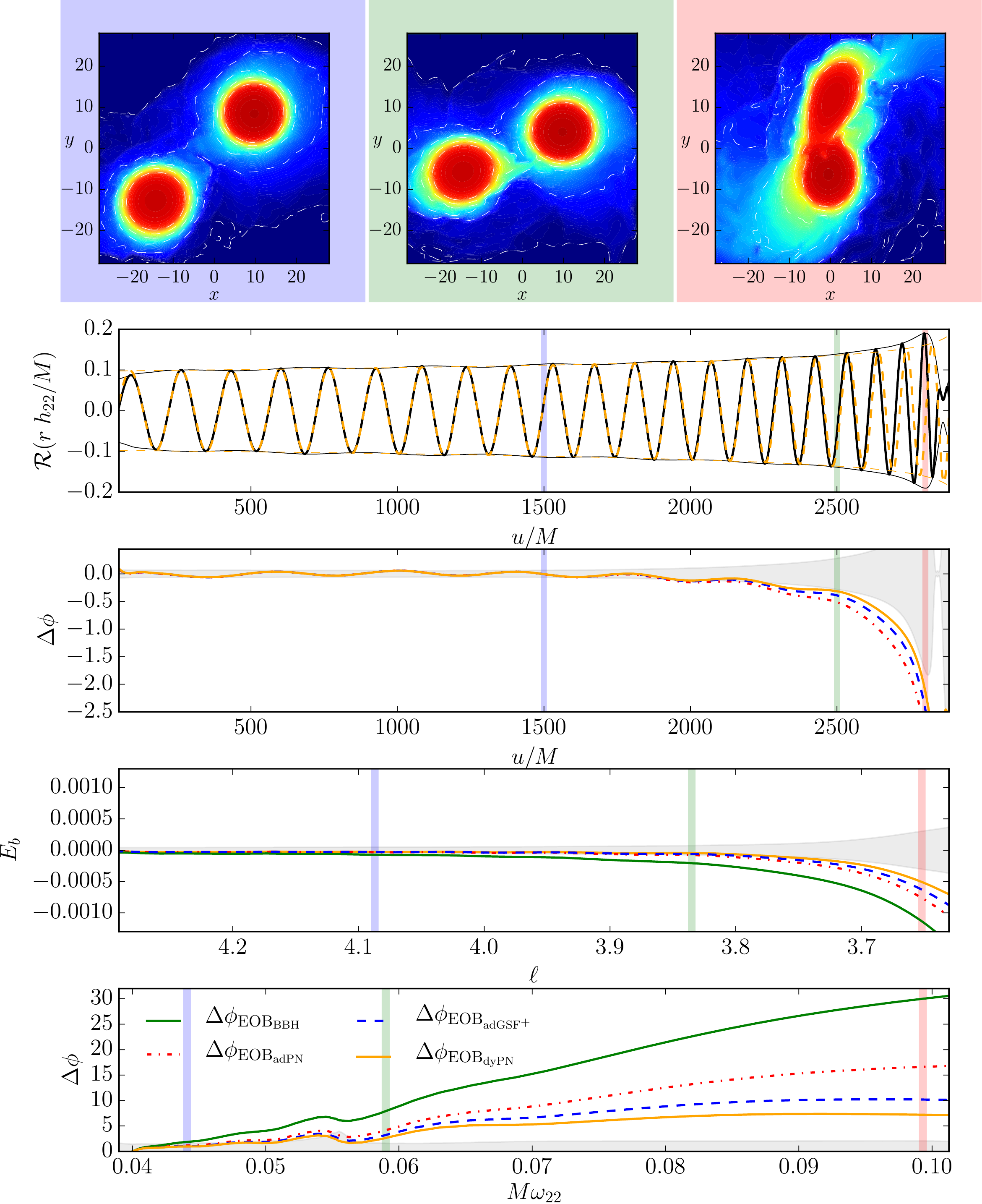}
   \caption{The top row shows rest-mass density snapshots with 
            the color bar ranging from $10^{-9}$ (dark blue)
            to $10^{-2.8}$ (dark red) for MS1b-165110. 
            White dashed lines show contour densities lines with
            $10^{-9},10^{-8},10^{-7},10^{-6},10^{-5},10^{-4},10^{-3}$. 
            The snapshots correspond to times $t=1500M$ (blue),
            $2500M$ (green) and the moment of merger $2802M$ (red).
            The second row shows the real part and the amplitude of the 
            GW for the NR data (black) and \EOBdy (orange). 
            The third row shows the the dephasing of the EOB models with 
            respect to the NR simulation. 
            The fourth row shows the difference of the binding energy 
            between the EOB models (including green \EOBpp) and the NR simulation. 
            The fifth row shows the dephasing as a function of the frequency.
            In all panels we use the same color scheme for all models. 
            The times corresponding to the snapshots are marked as shaded regions. }
   \label{fig:case:MS1b-165110} 
 \end{figure*}

Figure~\ref{fig:case:MS1b-165110} illustrates the results for the system MS1b-165110 
in a similar way as Fig.~\ref{fig:case:SLy-138138} does for 
SLy-138138. 

At time $t=1500M$ (left density snapshot) both stars are 
clearly separated, however, because of the small compactness of the lower-mass star 
and a stiff EOS, tidal effects are already visible at 
this early stage of the simulation.
Furthermore, the lower density regions are more affected by the artificial atmosphere, 
showing that the simulation of MS1b-165110 is less accurate 
than SLy-138138 and leading to larger NR uncertainties. 
We find that at $t=1500M$ all EOB models are 
able to describe the dynamics and GW emission of this system. 

Up to a time  $t \approx 2300 M $ all tidal EOB models
remain within the NR error estimate and hence consistent with the simulation. 
At this time the NSs are still separated. Some of the low density material in the NR 
simulation is transferred from the secondary to the more massive NS, however, 
as shown in~\cite{Dietrich:2016hky} this mass transfer is in part a numerical artifact that decreases with 
increasing resolution. Later, at $t=2500M$ or about two orbits before the merger 
(middle density snapshot) tidal deformations are significant and the EOB models deviate from the NR models.

Around merger the lower-mass star is tidally disrupted 
by the primary star (right density snapshot). 
Tidal effects are significantly larger than for SLy-138138. 
The dephasing caused by tidal effects ($\Delta \phi_{{\rm EOB}_{\rm BBH}}(\omega)$)
is of the order of 30 radiants, i.e.~more than a factor of two larger 
than for SLy-138138 (although the system merges at a smaller GW frequency).  
Overall, during the last four GW cycles before the merger none of the EOB models 
give a representation that is in agreement with the NR uncertainty. 
Neither the enhanced tidal effects due to the GSF$^+$-potential 
nor the dynamical tides of \EOBdy are strong enough for an accurate modeling, 
although again the \EOBdy performs better than the \EOBre model. 
Our results suggest that an additional enhancement of tidal 
effects in the EOB description for 
such regions of the BNS parameter space is needed. 
 
\subsection{Towards improved EOB models}
\label{sec:Paul}

The studies presented in this paper20170207 show that for the region 
of the BNS parameter space and length of NR simulations considered here
both the \EOBdy and \EOBre tidal EOB models provide a robust 
description of BNSs during most of the inspiral, 
however, our NR simulations reveal weaknesses of the models in the last few orbits before the merger. 
In particular, for cases where tidal effects are large
i.e.~for large $\kappa_2^T$,   
the EOB models tend to underestimate tidal effects 
irrespective of the mass ratio.
To quantify the impact of these differences 
on measurements with advanced GW detectors 
requires parameter estimation studies 
that are the subject of ongoing work, 
see also~\cite{Lackey:2013axa,Wade:2014vqa,Hotokezaka:2016bzh}. 
Nevertheless, understanding and improving 
waveform models in the regime close to merger 
will be essential for future GW detectors. 
Thus, we want to mention three possibilities for advancing existing models
in this part of the BNS parameter space.

\subsubsection{ Combining Dynamical tides with the GSF-inspired potential: \EOBPaul }

\begin{figure}[t]
   \includegraphics[width=0.48\textwidth]{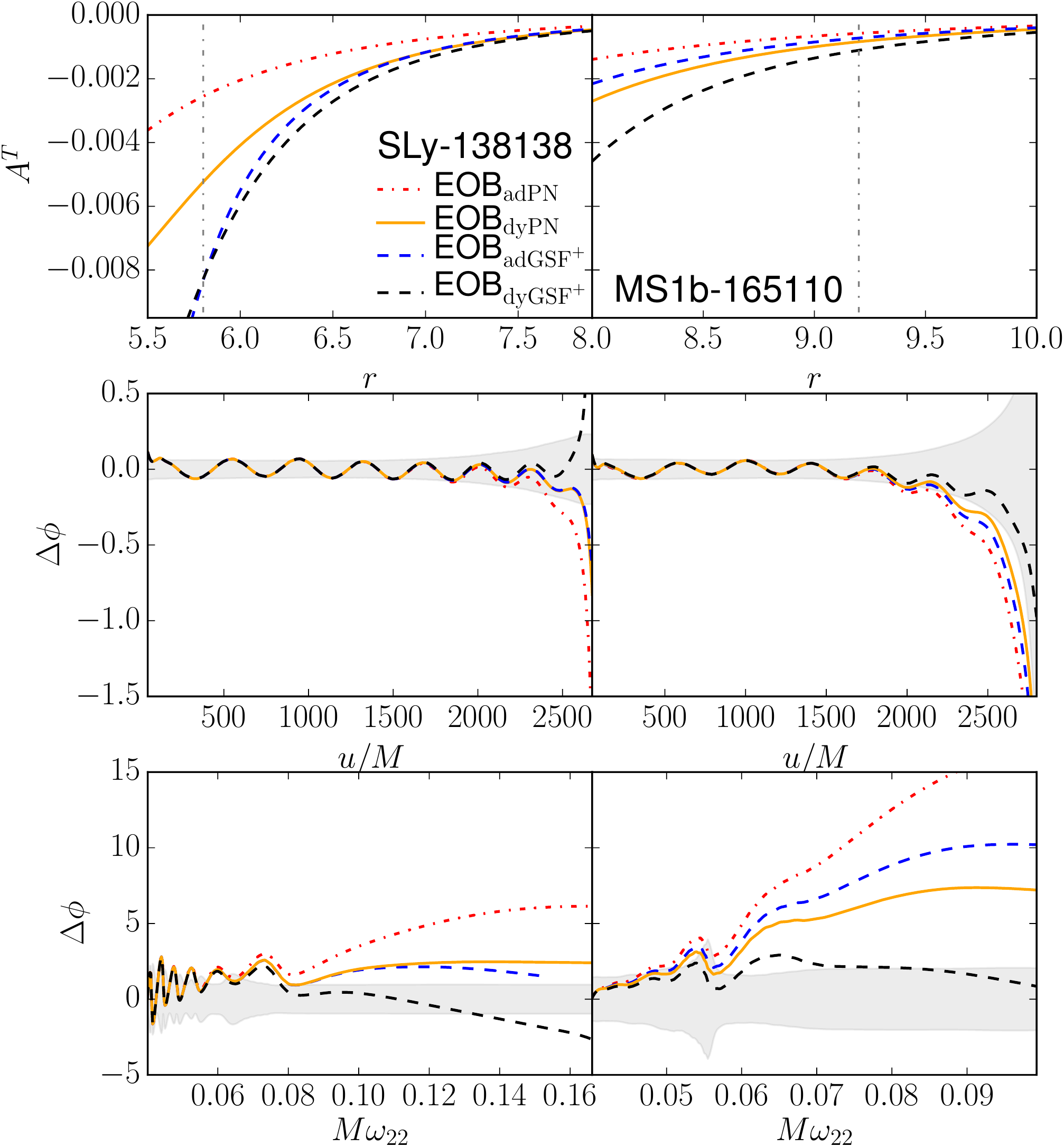}
   \caption{Top: Tidal contribution to the EOB-$A$-potential: $A^T$
            for SLy-138138 (left) and MS1b-165110 (right).
            The vertical dashed-dotted line marks the radius $r$ of the \EOBdy model 
            at the NR merger time.
            Middle: Dephasing $\Delta \phi(t)$ between the EOB models and 
            NR waveform. 
            Bottom panels: Dephasing $\Delta \phi(\omega)$ between the EOB models and 
            NR waveform.
            }
   \label{fig:EOBPaul} 
 \end{figure} 

As discussed above, the current description of tidal effects in the \EOBdy model tends to 
underestimate tidal effects in the late inspiral. 
To illustrate the effect of a simple enhancement of tidal 
contributions near the merger, we exchange the 
PN $A$-Potential with the GSF 
inspired potential introduced in 
Sec.~\ref{sec:tidalEOB:LR}.
This leads to a combination of \EOBre and \EOBdy: \EOBPaul. For this combination, 
the light ring $r_{\rm LR}$ is computed from Eq.~\eqref{eq:rLRdet} but
with the potential $\tilde{A}=A_{\rm pp}+A^{\rm tidal}_{\rm dyPN}$. 
While \EOBPaul allows a more accurate 
representation of systems with large $\kappa_2^T$ 
it also enhances tidal effects for systems with 
small $\kappa_2^T$ and leads to problems. 
The effect of using the combination \EOBPaul is presented 
in Fig.~\ref{fig:EOBPaul}, where the tidal part of the EOB 
$A$-potential is shown atop the phasing plots. The configurations are
SLy-138138 (left panels) and MS1b-165110 (right panels), 
i.e.~the two cases discussed in detail before.

As seen from the right panels, the
\EOBPaul model has a smaller dephasing from the NR results 
for systems with large $\kappa_2^T$ than the other models. 
In fact it is only for SLy-138138 that the dephasing 
of \EOBPaul is comparable to \EOBre and \EOBdy, while
for all other setups \EOBPaul represents the NR 
data better than the current models (not shown here). 

However, overestimating tidal effects for SLy-138138 is 
a substantial weakness of this ad-hoc 
combination of dynamical tides and the GSF inspired potential. While several possibilities to 
overcome these issues exist, a preferable approach that we will pursue in the future is to 
first understand and incorporate additional physical effects.

\subsubsection{Including additional physical effects}

As demonstrated by the detailed analyses in this paper20170207, improvements to tidal EOB models 
for the late inspiral will require including additional physical effects that are missing from the current 
descriptions. For example, existing tidal EOB models assume that the NSs are described as 
separated objects up to the merger. We showed in Fig.~\ref{fig:2d:merger}, Fig.~\ref{fig:case:SLy-138138}, and 
Fig.~\ref{fig:case:MS1b-165110} that the
low density material of the two NSs comes into contact already 
well before the merger. Subsequently, yet still prior to the merger, the tidally deformed
high-density cores of the NSs are moving inside the viscous material 
from the former NS outer layers. These physical effects must be taken into account in the tidal EOB models. 
However, a caveat with this conclusion is the fact that with increasing NR resolution and 
improved numerical methods the NS's surface becomes less `smeared out'
and the interaction of the low density material decreases. Therefore, we postpone a detailed 
modeling of these near-merger effects to future work, 
where we will consider NR simulations with even higher resolution than presented here. 

\subsubsection{Calibrating to NR waveforms}

Finally, an improvement of the tidal EOB models 
could also be achieved by incorporating information from NR simulations as done for 
EOB models for BBH coalescences, see e.g.~\cite{Taracchini:2013rva,Damour:2014sva,Bohe:2016gbl}.  
Calibrating to NR data entails introducing additional terms into the models and determining their 
coefficients by maximizing the overlap between the EOB and NR waveforms. 
This approach will require a large set of highly accurate NR 
simulations and possibly eccentricity reduced data. 
After calibration the models 
\textit{have to} be checked against a variety of 
waveforms covering a large region in the parameter 
space. This is a challenging task since the unknown EOS and the 
larger computational cost for general relativistic
hydrodynamic simulations make a full coverage 
of the BNS parameter space substantially more 
difficult than for BBHs. 

\section{Conclusion}
\label{sec:conclusion}
 
In this article we have presented the largest study to date 
of the performance of tidal EOB models for BNS systems. We considered setups covering a large 
portion of the BNS parameter space and analyzed variations in the total mass, the mass ratio, and the EOS. 
Furthermore, we performed a variety of tests to assess the successes and shortcomings of the 
models by considering several diagnostic quantities, some of which had
not been studied in the context of BNSs before. 

Our setups encompassed eight different configurations with two different 
EOSs, three different mass ratios, and two different total 
masses. We presented comparisons of the energetics of the BNS coalescence 
characterized by the binding energy vs.~specific angular momentum for all models, 
the periastron advance for one configuration,
the time evolution of the GW phase $\phi(t)$ of the 
(2,2), (2,1), (3,3), and (4,4) 
modes, and the frequency evolution of the GW phase 
$\phi(\omega)$ of the (2,2)-mode.

The most effective tool to study the GW signal emitted during the late stage of the 
BNS coalescence was to compare the accumulated 
phase within a given frequency interval $\Delta \phi(\omega)$. Considering this quantity enabled 
studies of the GW phasing without having to align waveforms and emphasized details of the last few orbits 
before the merger, where tidal effects are strongest. 

Overall we found that the \EOBdy model currently gives the best representation of 
the dynamics and GWs in the late-inspiral phase, however, the performance of the \EOBre 
model is comparable and differences between the models are small. 
For all setups the \EOBad model showed a significantly larger deviation from the 
NR results. We also found that the mass ratio has only a small influence 
on the performance of the tidal EOB models. Comparing results for higher modes led to similar 
results as for the dominant (2,2)-mode. We concluded that the main limitation of current tidal 
EOB models is that tidal effects are underestimated 
for systems with large tidal coupling constants, i.e.~those for which tidal effects are largest. 
This is a characteristic feature of all tidal EOB models which we have tested. 

In addition to comparisons between existing tidal EOB models and 
new state-of-the-art NR simulations, we also outlined possible avenues towards an improved 
waveform modeling. As a simple extension of the \EOBdy and the \EOBre models we considered the 
combination of both, i.e.~the use of dynamical tides and a GSF inspired potential as a 
simple way to mimic the larger tidal effects near the merger due to missing physics in the models. 
Although this method is `ad hoc' and not motivated by the inclusion of additional 
physical effects, the \EOBPaul model gives a phase evolution $\phi(t)$ closer 
to the NR results for almost all cases considered. 

We emphasize that besides improving
the non-spinning tidal EOB model, an important goal for future 
work on EOB tidal models is to fully include the effects of spins 
to achieve a better representation of the inspiral dynamics over a larger region 
of the BNS parameter space. 
While first NR simulations have characterized the effect 
of the NS spin on the GW phasing~\cite{Dietrich:2016lyp}, 
higher accuracy is needed to place constraints on EOB models, 
where spin effects for point masses are already incorporated.  
Further, the extension of the EOB models beyond the moment of merger 
is of importance and would improve their usability. 
While first attempts for a post-merger waveform model 
have been made, e.g.~\cite{Clark:2015zxa}, it 
may also be possible to derive a post-merger model 
using the same parameters characterizing the 
inspiral~\cite{Bernuzzi:2015rla}. 

Finally, we note that determining the priorities for future work on BNS systems will require inputs 
from data analysis studies. While it is desirable to have an EOB model that exactly reproduces
accurate NR results for the GW phasing and energetics up to the merger, 
it is important to assess the extent to which the discrepancies for 
some of the setups discussed in this paper20170207 would contaminate 
measurements with LIGO and other ground-based GW detectors. 
6

\begin{acknowledgments}
  It is a pleasure to thank 
  S.~Bernuzzi, B.~Br\"ugmann, A.~Buonanno, B.~Lackey, A.~Nagar, J.~Steinhoff, A.~Taracchini, 
  W.~Tichy, M.~Ujevic for helpful discussion and useful comments on the manuscript. 
  We also want to thank W.~Tichy for developing and sharing his SGRID code,
  S.~Bernuzzi and B.~Br\"ugmann for allowing us to use the BAM code, 
  A.~Taracchini for help with the EOB implementation, and  
  J.~Steinhoff for computing the $f-$mode frequencies.
  Computations where performed on SuperMUC at the LRZ (Munich) under 
  the project number pr48pu, Jureca (J\"ulich) 
  under the project number HPO21, Stampede 
  (Texas, XSEDE allocation - TG-PHY140019), 
  and the compute cluster Minerva of the Max Planck Institute for 
  Gravitational Physics (Albert Einstein Institute).
\end{acknowledgments}

\appendix

\section{Eccentricity from NR data}
\label{app:NRecc}

\subsection{Eccentricity from the orbital frequency} 

\begin{figure}[t]
   \includegraphics[width=0.5\textwidth]{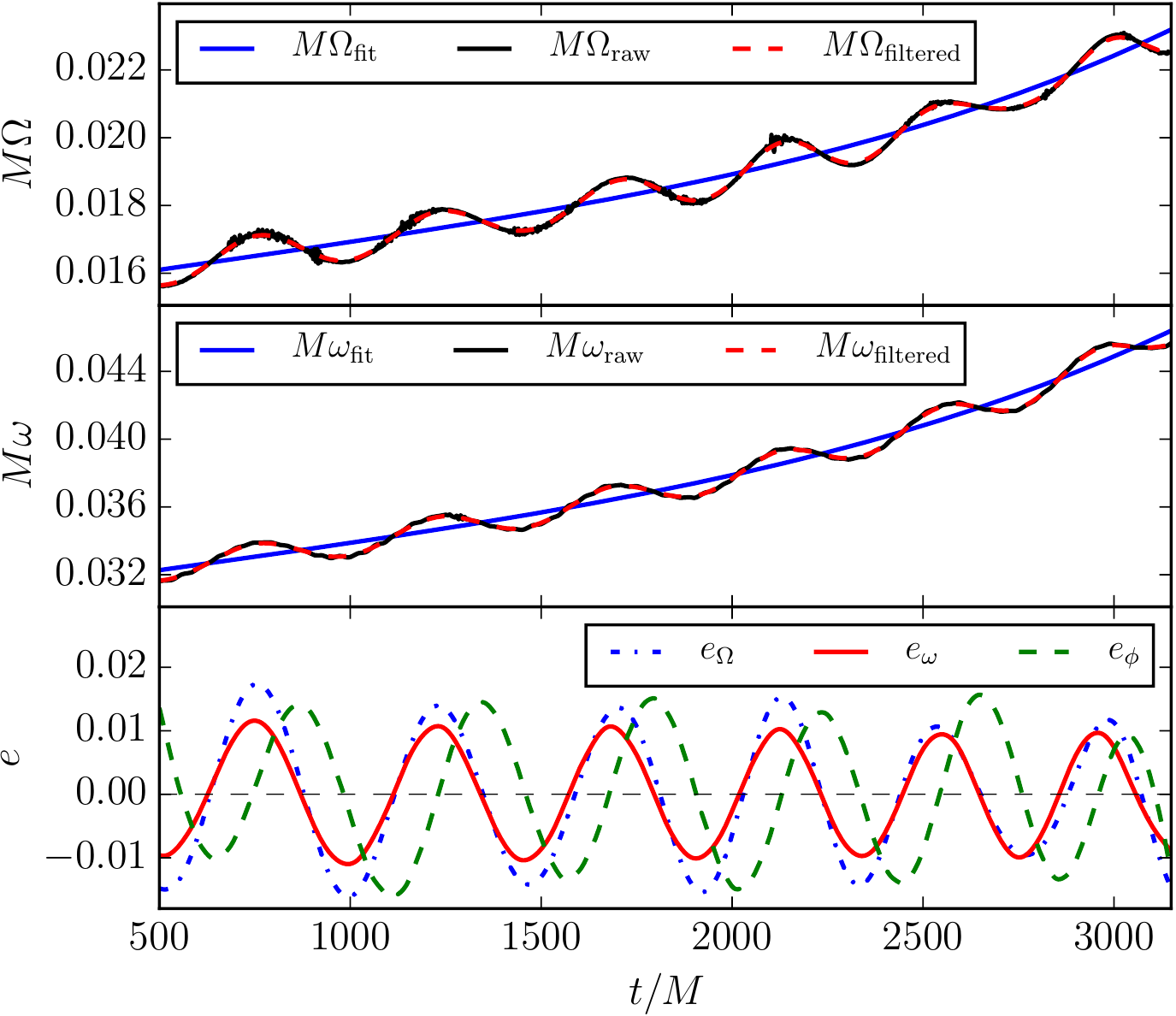}
   \caption{Eccentricity measures for MS1b-150100. 
            Upper panel: dimensionless orbital frequency $M \Omega$ 
            including the raw NR data (black solid line), 
            low-pass filtered NR data (red dashed line), 
            fitted data according to Eq.~\eqref{eq:omega_fit} (blue solid line). 
            Middle panel: dimensionless GW frequency $M \omega$ 
            including the raw NR data (black solid line), 
            low-pass filtered NR data (red dashed line), 
            fitted data according to Eq.~\eqref{eq:omega_fit} (blue solid line). 
            Lower panel: eccentricity measurements from the orbital frequency (blue dash-dotted)
            GW frequency (red solid) and the GW phase (green dashed). 
            }
   \label{fig:ecc_NR} 
 \end{figure}

The orbital frequency can be approximated from the 
coordinate location of the two NSs by 
\begin{equation}
 \Omega(t) = \frac{|\vec{r}(t) \times \dot{\vec{r}}(t)|}{|\vec{r}(t)|^2}, 
\end{equation}
where $\vec{r}(t)$ is the coordinate distance between the stars' centers. 
(The center of each star is defined as the minimum of the lapse function inside the star.) 
Because of the non-zero eccentricity and numerical noise $\Omega(t)$ oscillates during the inspiral. 
We remove high-frequency noise with a low-pass filter. 
To extract the eccentricity we fit the orbital frequency according to 
\begin{align}
  \Omega_{\rm{fit}} = \frac{1}{4}\tau^{-3/8}\left(1+c_1 \tau^{-1/4}+c_2 \tau ^{-3/8}\right) \nonumber \\
  + (a_1 t + b_1)^2 + (a_2 t + b_2)^4, 
  \label{eq:omega_fit}
\end{align}
with 
\begin{equation}
  \tau^2 = \frac{\nu^2(t_c-t)^2}{25M^2}+\tau_0^2,
\end{equation}
where $a_1$, $a_2$, $b_1$, $b_2$, $c_1$, $c_2$, $t_c$, and $\tau_0$ are fitting parameters. 
Eq.~\eqref{eq:omega_fit} is based on post-Newtonian calculations, e.g.~\cite{Hannam:2010ec},
but extended by $(a_1 t + b_1)^2 + (a_2 t + b_2)^4$. We find that by adding those terms 
the final fit is more robust than the choice of~\cite{Hannam:2010ec}, 
i.e.~less dependent on the fitting interval and the initial guess of the parameters. 
In Figure~\ref{fig:ecc_NR} (upper panel) we show the time evolution of 
$\Omega(t)$ for MS1b-150100. 
The raw data are presented in black. 
The eccentricity induced oscillations 
and smaller high frequency oscillations are clearly visible. We apply a low-pass 
filter to eliminate the high frequency oscillations (red dashed line)
and fit the filtered results with the model function $\Omega_{\rm fit}(t)$ (blue solid line). 
The fit depends on the fitting interval and needs to be chosen in a way that the 
initial junk radiation and gauge dynamics is cut out 
and that the signal does not extend too close to 
merger where $\Omega_{\rm fit}(t)$ loses validity. 
In Fig.~\ref{fig:ecc_NR} we restricted the fit and the analysis 
to times $t\in [500M,3100M]$. 
The final eccentricity is then given as
\begin{equation}
e_{\Omega}  =   \frac{\Omega(t) - \Omega_{\rm fit}(t)}{2 \Omega_{\rm fit }(t)}  \label{eq:eOmega}
\end{equation}
and shown for MS1b-150100 as a blue dash-dotted line in Fig.~\ref{fig:ecc_NR}.
Due to the emission of GWs the 
eccentricity decreases over time which can be seen by an increasing amplitude of 
$e_{\rm \Omega}$.

\subsection{Eccentricity from the GW frequency} 
 
In addition to the coordinate location of the two NSs we also use the GW signal 
to determine the eccentricity. For this purpose we compute the dimensionless GW frequency
$M \omega_{22}$.
[In the following part of this section, 
we drop the indices $_{22}$ for the (2,2)-mode for simplicity.] 
We proceed as for $\Omega(t)$ and apply a low-pass filter. 
Finally, we fit  $M \omega_{\rm filtered}$
according to Eq.~\eqref{eq:omega_fit} (blue line), 
see middle panel of Fig.~\ref{fig:ecc_NR}.
The eccentricity is then given as 
\begin{equation}
 e_{\omega}  =   \frac{\omega(u) - \omega_{\rm fit}(u)}{2 \omega_{\rm fit }(u)}. \label{eq:eomega} 
\end{equation}
We present $e_{\omega}$ for MS1b-150100 in the bottom panel of Fig.~\ref{fig:ecc_NR}
as a solid red line. 

\subsection{Eccentricity from the GW phase}

The last eccentricity indicator is based on the phase $\phi$ of the (2,2)-mode. 
To extract the eccentricity from the phase evolution, we fit $\phi$ according to 
\begin{equation}
\phi_{\rm fit} = \sum_{i=0}^{6} a_i t^i 
\end{equation}
after applying a low-pass filter. 
The eccentricity is then given as
\begin{equation}
 e_{\phi}    =   \frac{\phi(u) - \phi_{\rm fit}(u)}{4}. \label{eq:ephi}
\end{equation}
The eccentricity $e_{\phi}$ for MS1b-150100 is shown in the bottom panel of Fig.~\ref{fig:ecc_NR}
as a green dashed line. 

\section{Extracting $\phi(\omega)$ from NR data}
\label{app:phiomega}
In previous work we used a non-linear fit of the 
GW phase to compute the phase evolution as a function of time~\cite{Dietrich:2016lyp}. 
This approach is similar to that used for the PA in Appendix~\ref{app:NRecc}, 
but more delicate since the fit has to extend up to the merger. 
To overcome this problem, we use a different method in this work.

\begin{figure}[t]
   \includegraphics[width=0.5\textwidth]{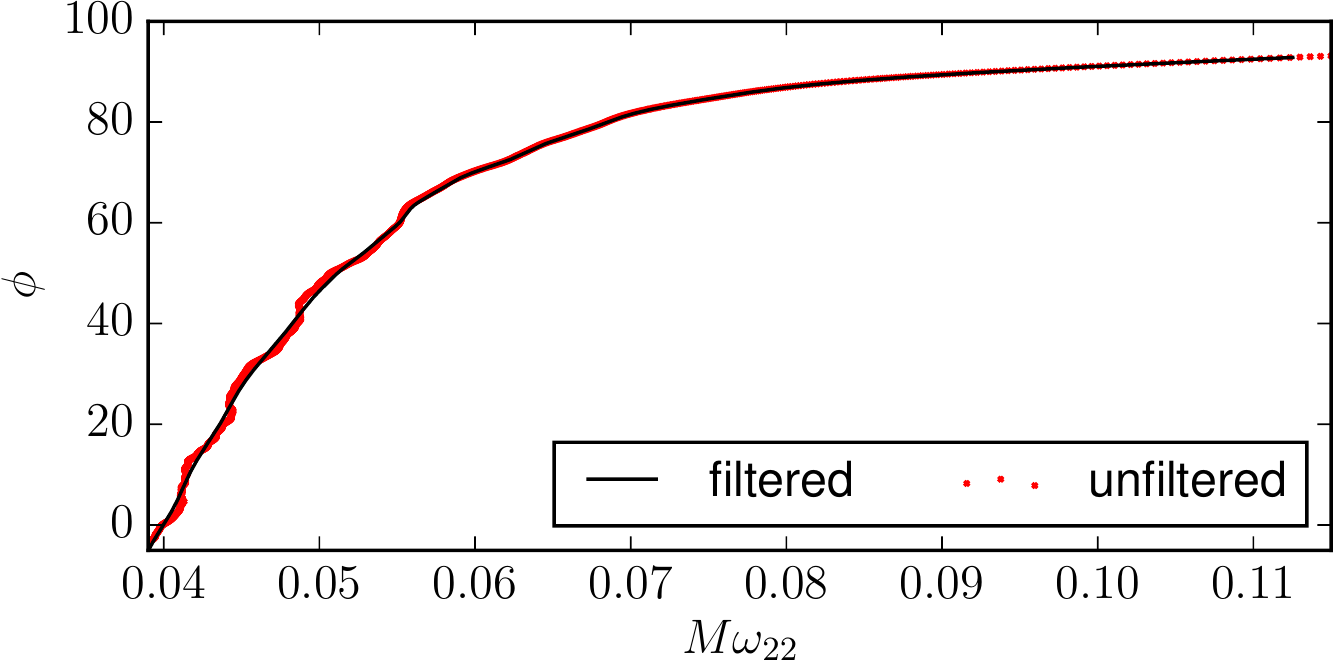}
   \caption{Computation of $\phi(\omega)$ for MS1b-153122. 
   We present the unfiltered NR data as small red crosses 
   and the interpolated and low pass filtered result as a solid black line. 
   The filtering removes most, but not all, of the eccentricity and 
   numerical noise.}
   \label{fig:phiomega_MS1b_153122} 
 \end{figure}

The biggest problem is that due to eccentricity and numerical noise 
$\omega(t)$ might be non-monotonic. 
We minimize the residual eccentricity with a low-pass 
filter applied to $\phi(t)$ and $\omega(t)$. 
Since the eccentricity decreases during the evolution while
the phase and frequency evolution accelerates, 
it is not possible to apply a low pass filter with a fixed 
cutoff frequency. 
Therefore we evaluate two different filtered functions and perform a smooth 
transition between them centered around $M\omega_{22}=0.055$ 
(except for MS1b-150100, where we use $M\omega_{22}=0.053$ since the signal is shorter). 
As an exemplary case we show the setup MS1b-153122 
in Fig.~\ref{fig:phiomega_MS1b_153122}, 
where the raw data are shown as red crosses and the fit as a 
black line. 
Notice that the NR data cluster at low frequencies since they are equally spaced in time.
Although the filtering reduces the eccentricity, some 
remaining eccentricity in particular at the transition 
between the two filtered function can be present. 
This is the reason for oscillations of $\Delta \phi(\omega)$ 
in the lower panels of Fig.~\ref{fig:phiomega}. 

\section{Influence of the underlying point mass EOB framework}
\label{app:EOBpp}

\begin{figure}[t]
   \includegraphics[width=0.5\textwidth]{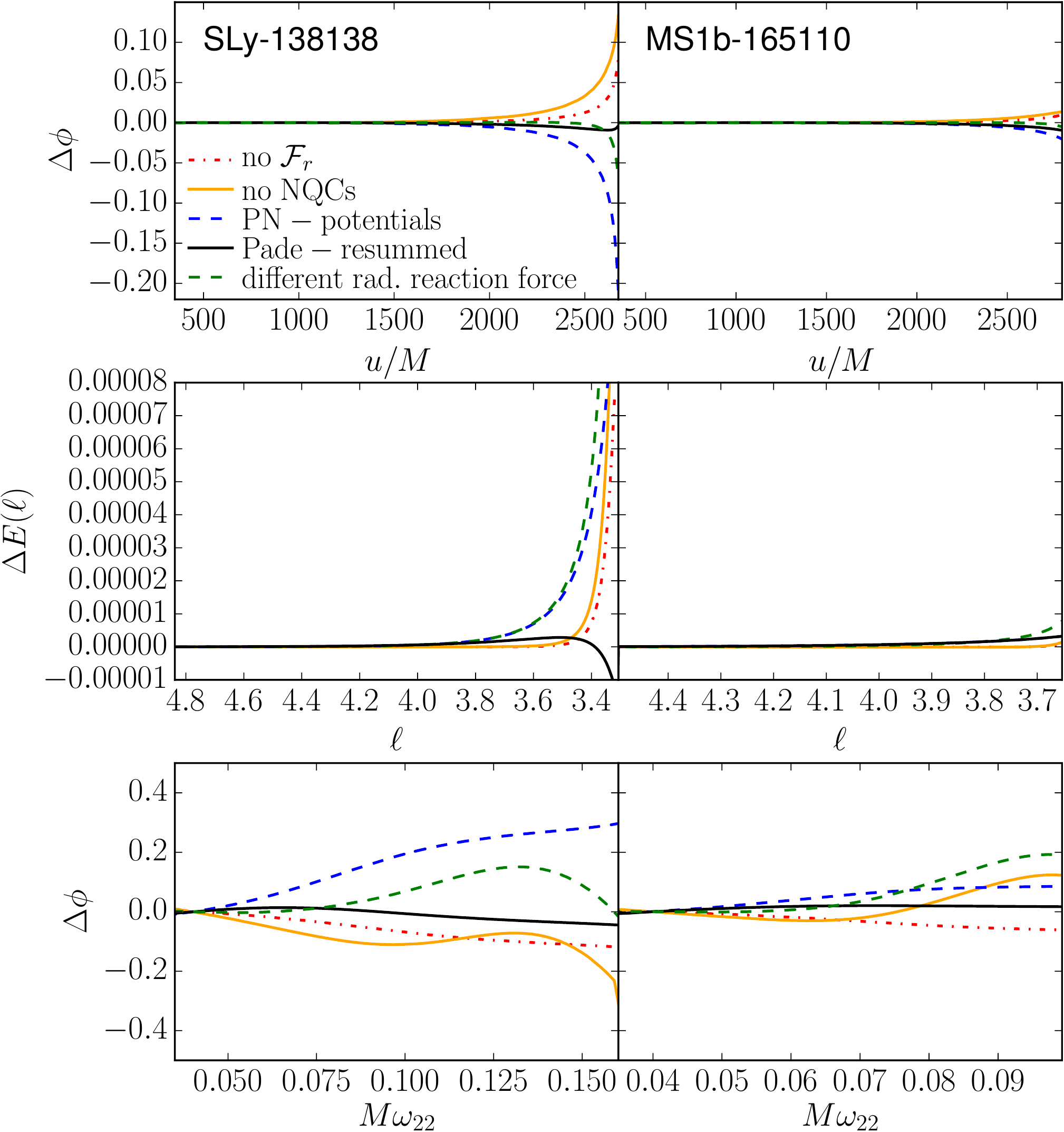}
   \caption{Influence of the details of the underlying point mass EOB model. 
   As a baseline we use the \EOBdy model as employed throughout the paper20170207.
   We study the effect of NQC correction, 
   the radial radiation reaction force, 
   and the explicit form of the EOB potentials, 
   see text for more details.  
   Top: Dephasing of the (2,2) mode as a function of time in radiants for the 
   SLy-138138 setup (left) and the MS1b-165110 setup (right). 
   Middle: Difference in binding energy for $\phi(\omega)$ for 
   SLy-138138 setup (left) and the MS1b-165110 setup (right). 
   Bottom: Dephasing of the (2,2) mode in radiants as a function of dimensionless 
   GW frequency for the SLy-138138 setup (left) and the MS1b-165110 setup (right). 
   }
   \label{fig:EOBpp} 
 \end{figure} 

Although this paper20170207 focuses on the performance of tidal EOB models using different ways of 
including tidal effects, we also want to briefly quantify the influence of the 
underlying point mass EOB model during the merger of BNS systems. 
For this reason we present again results for SLy-138138 and MS1b-165110 
as in Sec.~\ref{sec:case}.
The particular effects we study are: 
\begin{enumerate}[(i)]
 \item the radial radiation reaction force $\mathcal{F}_r$,
       where we compare results with $\mathcal{F}_r=0$ as employed in~\cite{Damour:2012ky,Bernuzzi:2014owa}
       with those that use $\mathcal{F}_r = (p_r/p_\phi) {\cal F}_\phi$ as in~\cite{Taracchini:2012ig,Hinderer:2016eia}, (shown red in Fig.~\ref{fig:EOBpp}).
 \item NQC corrections using the prescription of~\cite{Taracchini:2012ig} 
       (shown orange in Fig.~\ref{fig:EOBpp}).
 \item calibration and resummation of the potentials, i.e.~using simply the 
       PN Taylor expanded EOB potentials (shown blue in Fig.~\ref{fig:EOBpp}). 
 \item different resummation techniques for the EOB potentials,
       where we compare a Pade resummation with the calibration summarized in~\cite{Bernuzzi:2014owa}
  with a logarithmic resummation of the EOB potentials as given explicitly in~\cite{Steinhoff:2016rfi} with the calibration from~\cite{Taracchini:2013rva} (shown black in Fig.~\ref{fig:EOBpp}).
 \item different arguments in the $h_{\ell m}^F$ modes and thus also the radiation reaction force, where we compare results from using, in some occurences, a different frequency-related variable that was advocated in~\cite{Damour:2008gu} and is obtained by replacing $\Omega \to \Omega^3/(\partial H_{\rm EOB}/\partial p_\phi)^2$ evaluated for circular orbits (shown green in Fig.~\ref{fig:EOBpp}).
\end{enumerate}

Figure~\ref{fig:EOBpp} summarizes our results. 
The baseline is the \EOBdy model as employed throughout the article. 
To quantify the effects (i)-(v), we compute the difference in the 
time evolution of the GW phase (top panels), 
the binding energy (middle panels), 
and the phase evolution 
as a function of the frequency (bottom panels). 

Clearly visible is that for all quantities the underlying point mass 
EOB model effects the SLy-138138 configuration 
more than the MS1b-165110 setup. 
This is caused by the larger compactness and smaller 
radii of the NSs for SLy-138138, consequently
the stars come closer and reach higher frequencies during the inspiral. 
Considering the differences 
$\Delta \phi(t) =  \phi_{\rm dyPN_{\mathcal{F}_r=0}} - \phi_{\rm dyPN}>0$ 
and 
$\Delta \phi(t) =  \phi_{\rm dyPN_{\rm \neg NQC}} - \phi_{\rm dyPN}>0$ 
we find that the omission of a radial reaction force or NQC corrections 
accelerates the inspiral, which mimics larger tidal effects. 
The opposite is true when using the PN-potentials. The particular resummation technique 
and argument of the $h_{\ell m}$ modes have only a minor effect
on $\phi(t)$.

Considering the binding energy curves, 
we find that except when using a Pade resummation, 
all other changes of the underlying point mass EOB model lead to 
$\Delta E(\ell) > 0$, i.e., 
the setups are less bound than the \EOBdy model used in this work. 

Finally, considering the influence of the underlying point mass model 
on $\phi(\omega)$ we find that all phase differences $\phi(\omega)$ are below 
$0.5 \rm rad$. This is significantly smaller than the dephasing accumulated by 
tidal effects and also significantly smaller than the difference between 
the \EOBdy model and the NR simulations 
(cf.~e.g.~bottom panel of Fig.~\ref{fig:EOBPaul}).


\bibliography{paper20170207.bbl}     

\end{document}